# EXAMINATION OF THE ASTROPHYSICAL *S*-FACTORS OF THE RADIATIVE PROTON CAPTURE ON $^2$H, $^6$Li, $^7$Li, $^{12}$C AND $^{13}$C

Sergey Dubovichenko, Albert Dzhazairov-Kakhramanov

*V. G. Fessenkov Astrophysical Institute "NCSRT" NSA RK*

*050020, Kamenskoe plato 23, Observatory, Almaty, Kazakhstan*

*dubovichenko@mail.ru, albert-j@yandex.ru*

Astrophysical *S*-factors of radiative capture reactions on light nuclei have been calculated in a two-cluster potential model, taking into account the separation of orbital states by the use of Young schemes. The local two-body potentials describing the interaction of the clusters were determined by fitting scattering data and properties of bound states. The many-body character of the problem is approximatively accounted for by Pauli forbidden states. An important feature of the approach is the consideration of the dependence of the interaction potential between the clusters on the orbital Young schemes, which determine the permutation symmetry of the nucleon system. Proton capture on $^2$H, $^6$Li, $^7$Li, $^{12}$C, and $^{13}$C was analyzed in this approach. Experimental data at low energies were described reasonably well when the phase shifts for cluster-cluster scattering, extracted from precise data, were used. This shows that decreasing the experimental error on differential elastic scattering cross sections of light nuclei at astrophysical energies is very important also to allow a more accurate phase shift analysis. A future increase in precision will allow more definite conclusions regarding the reaction mechanisms and astrophysical conditions of thermonuclear reactions.

## 1. Introduction

The explanation of chemical element formation in stars is one of the major topics of modern nuclear astrophysics. The nuclear physics approach to the origin of the elements explains the abundance of different elements by their nuclear characteristics under various physical conditions during their formation. In addition, nuclear reactions in astrophysics allow us to understand, for example, the luminosity of a star in different evolutionary stages. Thereby questions of nucleosynthesis are closely coupled, on one hand, to questions regarding structure and evolution of stars and the Universe and, on the other hand, to nuclear interaction properties.[1-3]

However, there are a number of complicated and still unsolved problems because of which a complete theory of formation and evolution of the objects in the Universe cannot be formulated now. Let us consider some examples of these unsolved problems directly related to nuclear astrophysics and nuclear interactions, which follow from the existing to date nuclear physics problems:[1] *The insufficiency of experimental data on the nuclear reaction cross-sections at low and ultralow (astrophysical) energies.*

This problem arises from the impossibility, at the current stage of the experimental science, to carry out direct measurements of the thermonuclear reactions cross-sections on earth for energies at which they proceed in stars. Further in the article we will expand on this problem, but now we are going to illustrate the main concepts and notions used for the description of the thermonuclear reactions.

The data on cross-sections or astrophysical *S*-factors of thermonuclear reactions, including radiative capture reactions and their analysis in the frame of different theoretical models, are the main source of information about the nuclear processes which take place in the Sun and stars. Consideration of such reactions is complicated as in many cases only theoretical predictions or extrapolations can make up for the lacking experimental information about characteristics of thermonuclear processes in stellar matter at ultralow energies.[4]

The astrophysical *S*-factor, which determines the reaction cross-section, is the main characteristic of any thermonuclear reaction – it shows the probability of a certain behavior of particles at vanishing energies. It can be determined experimentally, but in most cases only at the energies above 100 keV÷1 MeV, while the real astrophysical calculations, for example, the star evolution problem, require the values of the astrophysical *S*-factor at energies about 0.1-100 keV, corresponding to the temperatures in the star core about $10^6$ K÷$10^9$ K.

One of the methods for obtaining the astrophysical *S*-factor at zero energy, i.e. the energy of the order of 1 keV and less, is the extrapolation of experimentally determined values to the lower energy range.

The second and naturally most preferable method involves theoretical calculation of the *S*-factor of a thermonuclear reaction on the basis of certain nuclear models.[5] However, the analysis of all thermonuclear reactions in the frames of a unified theoretical point of view is quite a time-consuming problem and further we will consider only photonuclear processes with γ-quanta, namely the radiative capture reactions for some light nuclei.

The aim of using nuclear models and theoretical methods of calculation of thermonuclear reaction characteristics is the following: if a certain nuclear model describes correctly the experimental data of the astrophysical *S*-factor in the energy range for which the data exist, for example 100 keV÷1 MeV, then it is quite reasonable to think that this model will describe correctly the form of the *S*-factor at the lowest energies (about 1 keV), too.

This approach has advantages over a simple data extrapolation, because the model used has, as a rule, a certain microscopic justification from the standpoint of the general principles of nuclear physics and quantum mechanics.

As for the model choice, one of the models, which we use in present calculations, is the potential cluster model (PCM) of the light atomic nuclei with the classification according to the Young schemes. The model, in certain cases, contains the states forbidden for intercluster interactions (FS) and in the simplest form gives a lot of possibilities for making such calculations.[6]

The potential cluster model used here is constructed on the basis of an assumption that the considered nuclei have a two-cluster structure. The choice of the potential cluster model is determined by the fact that the probability of formation of nucleon associations (clusters) and the degree of their separation from each other are rather high. This is proven by multiple experimental data and theoretical calculations obtained over the last fifty years.[6,7] Thus, in many cases and for various light nuclei the one-channel potential cluster model is a good approximation to the situation, which really exists in atomic nuclei.

Let us outline the general scheme that leads to the real results in the calculations of the astrophysical *S*-factor of certain thermonuclear reactions with γ-quanta (in this case we consider the radiative capture reaction). For carrying out such calculations, it is necessary to have certain data and take the following steps:

1. Have at one's own disposition the experimental data of the differential cross-sections or excitation functions $\sigma_{exp}$ for the elastic scattering of the nuclear particles under consideration (for example – p and $^2$H) at lowest energies known at the present moment.

2. Carry out the phase shift analysis of these data or have the results of the phase shift analysis of similar data that were have done earlier, i.e. know the phase shifts $\delta_L(E)$ of the elastic scattering as a function of energy E. It is one of the major stages of the whole calculation procedure of the astrophysical S-factors in the PCM with FS, since it allows obtaining the potentials of the intercluster interaction at the next step.

3. Construct the interaction potentials V(r) (for example for the p$^2$H system) according to the obtained phase shifts of scattering. This procedure is called the potential description of the elastic scattering phase shifts in the PCM with FS and it is necessary to make it for the lowest energies.

4. When we have the intercluster potentials, it is possible to calculate the total cross-sections of the photodisintegration process, for example $^3$He($\gamma$, p)$^2$H, and the total cross-sections of the $^2$H(p, $\gamma$)$^3$He radiative capture process related to the former by the principle of detailed balance. As a result, you can calculate the total theoretical cross-sections $\sigma(E)$ of the photonuclear reactions.

5. Then, it is possible to calculate the astrophysical S-factor of the thermonuclear reaction, for example $^2$H(p, $\gamma$)$^3$He, i.e. S(E) at any lower energies.

Let us note that, as of today, the experimental measurements were made only for the astrophysical S-factor of the radiative proton capture on $^2$H down to 2.5 keV, i.e. in the energy range, which can be named as astrophysical. For all other nuclear systems, taking part in the thermonuclear processes, in the best case such measurements were thoroughly made only down to 50 keV, as, for example, for the p$^3$H system.

All of these steps can be schematically represented in the following form:

$$\sigma_{exp} \rightarrow \delta_L(E) \rightarrow V(r) \rightarrow \sigma(E) \rightarrow S(E).$$

This scheme is identical for all photonuclear reactions and is independent of the reaction energy or some other factors. It represents a general approach to the study of any thermonuclear reaction with $\gamma$-quanta, provided that it is analyzed in the frames of the potential cluster model with FS.

## 2. Calculation methods

### 2.1. *Cluster model*

The potential cluster model under consideration is very simple in application, since technically it is reduced to solving a two-body problem, which is equivalent to the problem of one body in the field of a central force. Therefore, an objection can be put forward that this model is absolutely inadequate to the many-body problem to which the problem of description of properties of the system consisting of A nucleons is related.

In this regard, it should be noted that one of the successful models in the theory of atomic nucleus is the model of nuclear shells (SM) that mathematically represents the problem of one body in the field of a central force. The physical grounds of the potential cluster model considered here go back to the shell model or, more precisely, there is a surprising connection between the shell model and the cluster model, which is mentioned in the literature as the nucleon association model (NAM).[6]

In the NAM and PCM, the wave function of a nucleus consisting of two clusters with the numbers of nucleons $A_1$ and $A_2$ ($A = A_1 + A_2$) has the form of an antisymmetrized product of totally antisymmetric internal wave functions of clusters $\Psi(1,\ldots A_1) = \Psi(R_1)$ and $\Psi(A_1+1,\ldots,A) = \Psi(R_2)$ multiplied by the wave function of their relative motion $\Phi(R = R_1 - R_2)$,

$$\Psi = \hat{A} \{\Psi(R_1)\Psi(R_2)\Phi(R)\}, \qquad (1)$$

where $\hat{A}$ is the operator of antisymmetrization under permutations of nucleons to different clusters. $R$ is the intercluster distance, $R_1$ and $R_2$ are the radius vectors of the position of clusters with respect to the center of mass.

Usually, cluster wave functions are chosen in such a way that they correspond to ground states of nuclei consisting of $A_1$ and $A_2$ nucleons. These shell wave functions are characterized by specific quantum numbers, including Young schemes $\{f\}$, which determine the permutation symmetry of the orbital part of the wave function of cluster relative motion.

The concept of Pauli-forbidden states,[6] for which total wave functions $\Psi$ with relative motion wave functions becoming zero for FS under antisymmetrization by all $A$ nucleons, is introduced in this cluster model. The ground state, i.e. the really existing for this potential bound state, of such cluster system is described, in the general case, by the wave function with nonzero number of nodes. We will use the Young scheme technique for WF node number determination, which we will describe below and which will be used for consideration of different cluster systems.

Thus, the idea of Pauli-forbidden states makes it possible to take into account the many-body character of the problem in terms of interaction potential between clusters.[6] In this case, in practice, the interaction potential is chosen in order to describe experimental data (scattering phase shifts) on the elastic scattering of clusters in the corresponding $L$ partial wave and preferably in the state with one particular Young scheme $\{f\}$ for the spatial part of the wave function of $A$ nucleons.

Since, the results of phase shift analysis in the limited energy range, as a rule, prevent unambiguous reconstruction of the interaction potential, an additional restriction on the potential is required to reproduce the binding energy of the nucleus in the corresponding cluster channel and some other static nuclear properties. In this case, cluster masses are identified with masses of corresponding free nuclei. This additional requirement, obviously, is an idealization, since it assumes that the nucleus is 100% clusterized in the ground state. Indeed, the success of this potential model in description of a system of $A$ nucleons in the bound state is determined by the actual degree of clusterization of the nucleus in the ground state.

In this model, the $NN$ interaction manifests itself, as in the shell model, by creating the mean nuclear field, and provides clusterization of the nucleus. The remaining "work" on formation of the necessary number of nodes of the wave function of cluster relative motion is executed by the Pauli principle. Therefore, it is expected that the domain of applicability of the considered model is limited by nuclei with pronounced cluster properties.

However, some nuclear characteristics of particular, even non-cluster, nuclei can be mainly determined by one specific cluster channel and small contribution of other possible cluster configurations. In this case, the applied single-channel cluster model makes it possible to identify the dominating cluster channel and separate those properties of the cluster system that are determined by this channel.

## 2.2. *Astrophysical S-factor*

The formula for the astrophysical *S*-factor of the radiative capture process is of the form[8]

$$S(EJ) = \sigma(EJ)E_{cm} \exp\left(\frac{31.335\, Z_1 Z_2\, \sqrt{\mu}}{\sqrt{E_{cm}}}\right), \quad (2)$$

where $\sigma$ is the total cross-section (barn), $E_{cm}$ is the center of mass energy of particles, $\mu$ is the reduced mass of particles in the initial channel (atomic mass unit) and $Z_{1,2}$ are the particle charges in units of elementary charge. The numerical coefficient 31.335 was obtained on the basis of up-to-date values of fundamental constants.[9]

The total radiative capture cross-sections for electric *EJ*(*L*) transitions in the cluster model are given, for example, in Refs. 10, 11 and are written as

$$\sigma(E) = \sum_{J,J_f} \sigma(EJ, J_f), \quad (3)$$

$$\sigma(EJ, J_f) = \frac{8\pi K e^2}{\hbar^2 q^3} \frac{\mu}{(2S_1+1)(2S_2+1)} \frac{J+1}{J[(2J+1)!!]^2} A_J^2(K) \sum_{L_i, J_i} \left|P_J(EJ, J_f) I_J\right|^2, \quad (4)$$

where

$$P_J^2(EJ, J_f) = \delta_{S_i S_f} (2J+1)(2L_i+1)(2J_i+1)(2J_f+1)(L_i 0 J 0 | L_f 0)^2 \begin{Bmatrix} L_i & S & J_i \\ J_f & J & L_f \end{Bmatrix}^2, \quad (5)$$

$$A_J(K) = K^J \mu^J \left(\frac{Z_1}{M_1^J} + (-1)^J \frac{Z_2}{M_2^J}\right), \quad (6)$$

$$I_J = \langle L_f J_f | R^J | L_i J_i \rangle. \quad (7)$$

Here, $q$ is the wave number in the initial channel particles; $L_f$, $L_i$, $J_f$, $J_i$ are the angular moments of particles in the initial and final channels; $S_1$, $S_2$ are the spins; $M_{1,2}$, $Z_{1,2}$, are the masses and charges of the particles in the initial channel, respectively; $K^J$, $J$ are the wave number and angular moment of γ-quantum in the final channel; $I_J$ is the integral over wave functions of the initial (*i*) and final (*f*) states as a function of relative cluster motion with the intercluster distance *R*. Sometimes, the spectroscopic factor $S_{Jf}$ of the final state is used in the given formulas for cross-sections, but it is equal to one in the potential cluster model that we used, as it is done in work Angulo *et al.*[10]

Using the formula for the magnetic transition $M1(S)$ caused by the spin part of the magnetic operator[12] we can obtain

$$P_1^2(M1) = \delta_{S_i S_f} \delta_{L_i L_f} S(S+1)(2S+1)(2J_i+1)(2J_f+1) \begin{Bmatrix} S & L & J_i \\ J_f & 1 & S \end{Bmatrix}^2, \quad (8)$$

$$A_1(M1,K) = i\frac{e\hbar}{m_0 c} K\sqrt{3}\left[\mu_1 \frac{m_2}{m} - \mu_2 \frac{m_1}{m}\right], \quad (9)$$

$$I_1 = \langle \Phi_f | \Phi_i \rangle, \quad (10)$$

where $m$ is the mass of the nucleus; $\mu_1$ and $\mu_2$ are the magnetic moments of the clusters taken from Ref. 13, and, for example, for $\mu_{2H} = 0.857\mu_0$ and $\mu_p = 2.793\mu_0$; $\mu_0$ is the nuclear magneton.

The expression in square brackets in Eq. (9) for $A_1(M1, K)$ has been obtained on the assumption that, in the general form, for the spin part of the magnetic operator,[14]

$$W_{Jm}(S) = i\frac{e\hbar}{m_0 c} K^J \sum_i \mu_i \hat{\vec{S}}_i \cdot \vec{\nabla}_i (r_i^J Y_{Jm}(\Omega_i)) \quad (11)$$

summation by $r_i$, i.e. by coordinates of center of mass of clusters, relative to common centre of mass of the nucleus, is held before the action of the $\nabla$-nabla operator which acts on the expression in brackets $(r_i^J Y_{Jm}(\Omega_i))$ and leads to decrease of $r_i$ degree,[12]

$$\vec{\nabla}_i (r_i^J Y_{Jm}(\Omega_i)) = \sqrt{J(2J+1)} r_i^{J-1} \vec{Y}_{Jm}^{J-1}(\Omega_i). \quad (12)$$

In this case the coordinates $r_i$ are $R_1 = m_2/mR$ and $R_2 = -m_1/mR$, where $R$ is the relative intercluster distance and $R_1$ and $R_2$ are the distances from the common center of mass to the centers of mass of each cluster.

The operator of electromagnetic transition for the interaction between radiation and matter, in electromagnetic processes like radiative capture and photodisintegration, is well known. Therefore, there is a fine possibility to clarify the form of two-particle strong interaction in the initial channel when they are in the continuous spectrum and bound states of the same particles in the final channel, i.e. in states of discrete spectrum.

### 2.3. *Potentials and wave functions*

The intercluster interaction potentials for each partial wave, i.e., for given orbital angular moment $L$, and point-like Coulomb term, were represented as (further, the only nuclear part of potential is given)

$$V(R)=V_0\exp(-\alpha R^2)+V_1\exp(-\gamma R) \quad (13)$$

or

$$V(R) = V_0 \exp(-\alpha R^2). \tag{14}$$

Here, $V_1$ and $V_0$ (dim.: MeV), $\alpha$ and $\gamma$ (dim.: fm$^{-2}$ and fm$^{-1}$, correspondingly). These are the potential parameters found from experimental data under the constraint of best description of the elastic scattering phase shifts extracted in the process of phase shift analysis from the experimental data of differential cross-sections, i.e. angular distributions or excitation functions.

The expansion of WF of relative cluster motion in nonorthogonal Gaussian basis and the independent variation of parameters[11] are used in the two-particle variational method (VM).

$$\Phi_L(R) = \frac{\chi_L(R)}{R} = R^L \sum_i C_i \exp(-\beta_i R^2), \tag{15}$$

where $\beta_i$ and $C_i$ are the variational expansion parameters and expansion coefficients.

The behavior of the wave function of bound states (BS) at large distances is characterized by the asymptotic constant $C_W$ of the form[15]

$$\chi_L = \sqrt{2k_0} C_W W_{-\eta L + 1/2}(2k_0 R), \tag{16}$$

where $\chi_L$ is the numerical wave function of the bound state obtained from the solution of the radial Schrödinger equation and normalized to unity; $W_{-\eta L+1/2}$ is the Whittaker function of the bound state determining the asymptotic behavior of the WF, which is the solution to the same equation without nuclear potential, i.e. long distance solution; $k_0$ is the wave number determined by the channel binding energy; $\eta$ is the Coulomb parameter; $L$ is the orbital moment of the bound state.

The mean square mass radius is represented as

$$R_m^2 = \frac{M_1}{M}\langle r_m^2 \rangle_1 + \frac{M_2}{M}\langle r_m^2 \rangle_2 + \frac{M_1 M_2}{M^2} I_2, \tag{17}$$

where $M_{1,2}$ and $\langle r_m^2 \rangle_{1,2}$ are the masses and square mass radii of clusters, $M=M_1+M_2$, $I_2$ - the integral of the form

$$I_2 = \langle \chi_L(R) | R^2 | \chi_L(R) \rangle, \tag{18}$$

for the inter cluster distance $R$ over radial wave functions $\chi_L(R)$ of relative cluster motion normalized to unity in the ground state of the nucleus with the orbital angular moment $L$. Form of this expression is similar to Eq. (7).

The mean square charge radius is represented as

$$R_z^2 = \frac{Z_1}{Z}\langle r_z^2 \rangle_1 + \frac{Z_2}{Z}\langle r_z^2 \rangle_2 + \frac{(Z_2 M_1^2 + Z_1 M_2^2)}{ZM^2} I_2, \tag{19}$$

where $Z_{1,2}$ and $\langle r_z^2 \rangle_{1,2}$ are the charges and square charge radii of clusters, $Z=Z_1+Z_2$, $I_2$ is the above mentioned integral.

The wave function $\chi_L(R)$ or $|L_i J_i\rangle$ is the solution of the radial Schrödinger equation of the form

$$\chi''_L(R) + [k^2 - V(R) - V_{\text{Coul.}}(R) - L(L+1)/R^2]\chi_L(R) = 0 ,\qquad(20)$$

where $V(R)$ is the inter-cluster potential represented as expressions (5) or (6) (dim. fm$^{-2}$); $V_{\text{Coul.}}(R)$ is the Coulomb potential; $k$ is the wave number determined by the energy $E$ of interaction particles $k^2=2\mu E/\hbar^2$.

Generally, all calculations are carried out by finite-difference method (FDM), which is the modification of methods[16] and take into account Coulomb interactions. The variational method with the expansion of the wave function in nonorthogonal Gaussian basis (see Eq. (15)) is used for an additional control of calculations of the binding energy and WF form.[11]

### 2.4. *Classification of cluster states*

The states with minimal spin, in the processes of scattering of some light atomic nuclei, turn out to be "mixed" according to Young orbital schemes, for example, the p$^2$H doublet state[17] is "mixed" according to schemes {3} and {21}. At the same time, the bound forms of these states, for example, the doublet p$^2$H channel of $^3$He is "pure" according to scheme {3}. The method of separation of such states according to Young schemes was proposed in Refs. 7, 17 where, in all cases, the "mixed" scattering phase shift can be represented as a half-sum of "pure" phase shifts $\{f_1\}$ and $\{f_2\}$

$$\delta^{\{f_1\}+\{f_2\}} = 1/2\left(\delta^{\{f_1\}} + \delta^{\{f_2\}}\right).\qquad(21)$$

In this case it is assumed that $\{f_1\}=\{21\}$ and $\{f_2\}=\{3\}$, and the doublet phase shifts, extracted from the experiments, are "mixed" in accordance with these two Young schemes. If we suppose that instead of the "pure" quartet phase shift with the symmetry {21}, one can use the "pure" doublet phase shift of p$^2$H scattering with {21} symmetry. Then it is easy to find the "pure" doublet p$^2$H phase shift with {3} symmetry[17] and use it for the construction of the "pure" interaction potential according to Young schemes that can be used for description of characteristics of the bound state.

Such potential allows us to consider the bound p$^2$H state of $^3$He. Similar ratios apply to other light nuclear systems as well, and further in each specific case we will analyze the AS and FS structure for scattering potentials of the ground bound states.[6]

### 2.5. *Phase shift analysis*

Using the experimental data of differential cross-sections of elastic scattering, it is possible to find a set of phase shifts $\delta^J_{S,L}$, which can reproduce the behavior of these cross-sections with certain accuracy. Quality of description of experimental data on the basis of a certain theoretical function (functional of several variables) can be estimated by the $\chi^2$ method, which is written as

$$\chi^2 = \frac{1}{N}\sum_{i=1}^{N}\left[\frac{\sigma_i^t(\theta)-\sigma_i^e(\theta)}{\Delta\sigma_i^e(\theta)}\right]^2 = \frac{1}{N}\sum_{i=1}^{N}\chi_i^2, \qquad (22)$$

where $\sigma^e$ and $\sigma^t$ are experimental and theoretical (i.e. calculated for some defined values of phase shifts $\delta_{S,L}^J$ of scattering) cross-sections of elastic scattering of nuclear particles for $i$-angle of scattering, $\Delta\sigma^e$ – the error of experimental cross-sections at these angles, $N$ – the number of measurements.[18]

The less $\chi^2$ value, the better description of experimental data on the basis of chosen phase shift of scattering set. Expressions, describing the differential cross-sections, represent the expansion of some functional $d\sigma(\theta)/d\Omega$ to the numerical series and it is necessary to find such variational parameters of expansion $\delta_L$, which are the best for description of its behavior. Since the expressions for the differential cross-sections are exact, then as $L$ approaches infinity, the value of $\chi^2$ must vanish to zero. This criterion is used by us for choosing a certain set of phase shifts ensuring the minimum of $\chi^2$ which could possibly be the global minimum of a multiparameter variational problem.[19,20]

The exact mass values of particles were taken for all of our calculations,[13] and the $\hbar^2/m_0$ constant was taken to be 41.4686 MeV fm$^2$. The Coulomb parameter $\eta=\mu Z_1 Z_2 e^2/(q\hbar^2)$ was represented as $\eta=3.44476\ 10^{-2} Z_1 Z_2\ \mu/q$, where $q$ is the wave number determined by the energy of interacting particles in the initial channel (in fm$^{-1}$), $\mu$ - the reduced mass of the particles (atomic mass unit), $Z$ - the particle charges in elementary charge units. The Coulomb potential with $R_{\text{Coul.}}=0$ was represented as $V_{\text{Coul.}}(\text{MeV})=1.439975\ Z_1 Z_2/r$, where $r$ is the distance between the initial channel particles (fm).

## 3. The radiative proton capture on $^2$H

The first process under our consideration is the radiative capture reaction

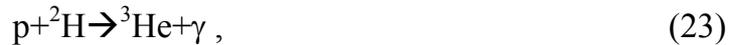
$$\text{p}+{}^2\text{H}\rightarrow{}^3\text{He}+\gamma, \qquad (23)$$

which is a part of proton-proton chain and gives a considerable contribution to energy efficiency of thermonuclear reactions[1] accounting for burning of the Sun and stars of our Universe. The potential barrier for interacting nuclear particles of the p-p chain is the lowest. Thus, it is the first chain of nuclear reactions which can take place at lowest energies and stellar temperatures.

For this chain, the radiative proton capture on $^2$H is the basic process for the transition from the primary proton fusion

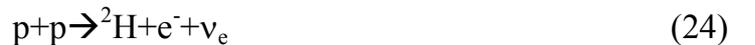
$$\text{p}+\text{p}\rightarrow{}^2\text{H}+e^-+\nu_e \qquad (24)$$

to the capture reaction of two $^3$He nuclei,[21] which is one of the final processes

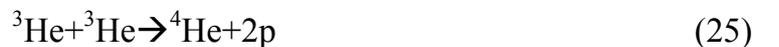
$$^3\text{He}+{}^3\text{He}\rightarrow{}^4\text{He}+2\text{p} \qquad (25)$$

in the p-p chain.

The theoretical and experimental study of the radiative proton capture on $^2$H,

in detail, is of fundamental interest not only for nuclear astrophysics, but also for nuclear physics of ultralow energies and lightest atomic nuclei.[22,23] That is why the experimental studies of this reaction are in progress and at the beginning of 2000[th] years the new experimental data in the range down to 2.5 keV appeared because of the LUNA European Project.

### 3.1. *Photoprocesses, potentials and scattering phase shifts*

The total cross-sections of the photoprocesses of lightest $^3$He and $^3$H nuclei were considered earlier, in the frame of the potential cluster model with forbidden states, in our work Ref. 24. The $E1$ transitions resulting from the orbital part of the electric operator $Q_{Jm}(L)$[11] were taken into account in these calculations for photodisintegration of $^3$He and $^3$H in the p$^2$H and n$^2$H channels. The cross-sections of $E2$ processes and cross-sections depending on the spin part of electric operator turned out to be several orders less.

Further, it was assumed that $E1$ electric transitions in the N$^2$H system are possible between ground "pure" (scheme {3}) $^2S$ state of $^3$H and $^3$He nuclei and doublet $^2P$ scattering state "mixed" according to Young schemes {3}+{21}.[22,23] Such transition is quite possible, since the quantum number, connected with Young schemes, evidently is not saved in electromagnetic processes.[24]

To calculate photonuclear processes in the systems under consideration the nuclear part of the potential of inter-cluster p$^2$H and n$^2$H interactions is represented as expression (5) with a point-like Coulomb potential, $V_0$ - the Gaussian attractive part, and $V_1$ - the exponential repulsive part. The potential of each partial wave was constructed in order to correctly describe the corresponding elastic scattering partial phase shift.[25-28] Using this concept; the potentials of the p$^2$H interaction of scattering processes were obtained. The parameters of these potentials are given in Refs. 11, 22-24, 29 and parameters for doublet states are listed in Table 1.

Table 1. The potentials of the p$^2$H interaction in doublet channel.[11,22-24]

| $^{2S+1}L$, {f} | $V_0$ (MeV) | α (fm$^{-2}$) | $V_1$ (MeV) | γ (fm$^{-1}$) |
|---|---|---|---|---|
| $^2S$, {3} | -34.76170133 | 0.15 | --- | --- |
| $^2P$, {3}+{21} | -10.0 | 0.16 | +0.6 | 0.1 |
| $^2S$, {3}+{21} | -55.0 | 0.2 | --- | --- |

With kind permission of the European Physical Journal (EPJ)

Then, in the doublet channel, mixed according to Young schemes {3} and {21},[17] "pure" phase shifts in Eq. (21) were separated, and based on these phase shifts the "pure" $^2S$ potential of the bound state with scheme {3} for $^3$He in the p$^2$H channel was constructed.[11,22-24,29]

The calculations of the $E1$ transition[24] have shown that the best results for description of total photodisintegration cross-sections for $^3$He in the γ-quantum energy range 6-28 MeV, including the maximum at $E_γ$=10-13 MeV, can be obtained if we use the potentials with peripheric repulsion of the $^2P$-wave for the p$^2$H scattering (see Table 1) and with the "pure" according to Young schemes $^2S$-interaction of the BS that has a Gaussian form (see Eq. (13)) with parameters

$$V_0 = -34.75 \text{ MeV}, \alpha = 0.15 \text{ fm}^{-2}, V_1 = 0, \quad (26)$$

which were obtained, primarily, on the basis of a correct description of the binding energy (with an accuracy of several kiloelectronvolts) and the charge radius of $^3$He. The calculations of the total cross-sections of the radiative proton capture on $^2$H and astrophysical $S$-factors were used with these potentials at the energy range down to 10 keV.[11,24] Though, at that period of time, we only knew $S$-factor experimental data in the range above 150-200 keV.[30]

Recently, the new experimental data on the $S$-factor of the proton capture on $^2$H at energies down to 2.5 keV appeared not long ago.[31-33] That is why it is interesting to know if it is possible to describe the new data on the basis of $E$1 and $M$1 transitions in the potential cluster model with the earlier obtained $^2P$-interaction of scattering and $^2S$-potential of the p$^2$H ground state (GS) of $^3$He.

The parameters of the "pure" doublet $^2S$-potential according to Young scheme {3} were adjusted for a more accurate description of the experimental binding energy of $^3$He in the p$^2$H channel. This potential (Table 1) has become somewhat deeper than the potential we used in Ref. 24 and leads to the total agreement between calculated -5.4934230 MeV and experimental -5.4934230 MeV binding energies obtained with exact values of particle masses.[13] The difference between the potentials given in Ref. 24 and in Table 1 is primarily due to the application of the exact masses of particles and more accurate description of binding energy of $^3$He in the p$^2$H channel. For all of these calculations, the absolute accuracy of calculation of the binding energy in our computer program based on the finite-difference method was taken to be at level of $10^{-8}$ MeV.

The charge radius of $^3$He with this potential equals 2.28 fm, which is a little higher than the experimental values listed in Table 2.[13,34,35] The experimental radii of proton and deuteron are used for these calculations and the latter is larger than the radius of $^3$He. Thus, if the deuteron is present in $^3$He as a cluster, it must be compressed by about 20-30% of its size in a free state for a correct description of the $^3$He charge radius.[11]

Table 2. Experimental masses and charge radii of light nuclei used in these calculations.[13,34,35]

| Nucleus | Radius (fm) | Mass |
|---|---|---|
| p | 0.8768(69) | 1.00727646677 |
| $^2$H | 2.1402(28) | 2.013553212724 |
| $^3$H | 1.63(3); 1.76(4); 1.81(5)<br>The average value is 1.73 | 3.0155007134 |
| $^3$He | 1.976(15); 1.93(3); 1.877(19); 1.935(30)<br>The average value is 1.93 | 3.0149322473 |
| $^4$He | 1.671(14) | 4.001506179127 |

The asymptotic constant $C_W$ with Whittaker asymptotics[36] given in Eq. (16) was calculated for controlling behavior of the WF of BS at large distances; its value in the range of 5-20 fm equals $C_W$=2.333(3). The error, given here, is found by averaging the constant in the range mentioned above. The experimental data known for this constant yield the values in an interval of 1.76-1.97,[37,38] which is slightly less than the value obtained here. It is possible to give results of three-body calculations,[39] where a good agreement with the experiment[40] for the ratio of asymptotic constants of $^2S$ and $^2D$ waves was obtained and the value of the constant of $^2S$ wave was found to be $C_W$=1.878.

However, in Plattner et al.,[15] which was published later than works Bornard et al.[37] and Platner et al.,[38] the value of 2.26(9) was given for the asymptotic constant, and this is in a good agreement with our calculations. It can be seen from the considerable data that there is a big difference between the experimental results of asymptotic constants obtained in different periods. These data are in the range from 1.76 to 2.35 with the average value of 2.06.

In the two-cluster model, the value of $C_W$ constant and charge radius strongly depend on the width of the potential well and it is always possible to find other parameters of $^2S$-potential of the bound state, for example:

$$V_0 = -48.04680730 \text{ MeV and } \alpha = 0.25 \text{ fm}^{-2}, \qquad (27)$$

$$V_0 = -41.55562462 \text{ MeV and } \alpha = 0.2 \text{ fm}^{-2}, \qquad (28)$$

$$V_0 = -31.20426327 \text{ MeV and } \alpha = 0.125 \text{ fm}^{-2}, \qquad (29)$$

which yield the same binding energy for $^3$He in the p$^2$H channel. The first of them at an interval of 5-20 fm leads to the asymptotic constant $C_W$=1.945(3) and the charge radius $R_{ch}$=2.18 fm, the second variant yields the constant $C_W$=2.095(5) and $R_{ch}$=2.22 fm, the third - $C_W$=2.519(3) and $R_{ch}$=2.33 fm.

It can be seen from these results that the potential in Eq. (27) allows one to obtain the charge radius that is closest to experimental value. Further, reduction of the potential width may result in correct description of its value; however, as it will be shown bellow, this fact will not make it possible to reproduce the S-factor of the proton capture on $^2$H. In this sense, the slightly wider potential Eq. (28) has the minimal acceptable width of the potential well, which leads to asymptotic constant practically equals its experimental average value 2.06 and acceptably describe the behavior of the astrophysical S-factor in the broadest energy region.

The variational method (VM)[20] was used for an additional control of the accuracy of binding energy calculations for the potential from Table 1, which allowed obtaining the binding energy of -5.4934228 MeV by using the independent variation of parameters and the grid having dimension 10. The asymptotic constant $C_W$ of the variational WF is on the level of 2.34(1) at distances of 5-20 fm in these calculations. The variational parameters and expansion coefficients of the radial wave function for this potential having form given in Eq. (15) are listed in Ref. 4.

The potential in Eq. (28) was examined within the frame of the VM and the same binding energy of -5.4934228 MeV has been obtained. The variational parameters and expansion coefficients of the radial wave function also are listed in Ref. 4. The asymptotic constant at distances of 5-20 fm turned out to be 2.09(1) and the residual error is of the order of $10^{-13}$.[20]

For the real binding energy in this potential it is possible to use the average value -5.4934229(1) MeV with the calculation error of finding energy by two methods equal to ±0.1 eV, because the variational energy decreases as the dimension of the basis increases and gives the upper limit of the true binding energy, but the finite-difference energy increases as the size of steps decreases and the number of steps increases.[20]

### 3.2. *Astrophysical S-factor*

In our present calculations of the astrophysical *S*-factor we considered the region of energies of the proton capture on $^2$H from 1 keV to 10 MeV and found the value of 0.165 eV b for the $S(E1)$-factor at 1 keV for the potentials from Table 1. The founded value is slightly lower than the known data if we consider the total *S*-factor without separation it into $S_s$ and $S_p$ parts resulting from *M*1 and *E*1 transitions. This separation was made in work Schimd, *et al.*,[32] where $S_s(0)$=0.109(10) eV b and $S_p(0)$=0.073(7) eV b. At the same time, the authors give the following values $S_0$=0.166(5) eV b and $S_1$=0.0071(4) eV b keV$^{-1}$ in the linear approximation formula

$$S(E_{c.m.}) = S_0 + E_{c.m.} S_1, \qquad (30)$$

and for *S*(0) leads to the value of 0.166(14) keV b, which was obtained with all possible errors. The results, taking into account the separation of the *S*-factor into *M*1 and *E*1 parts, are given in one of the first of works, Griffiths *et al.*,[30] where $S_s$=0.12(3) eV b, $S_p$=0.127(13) eV b for the total *S*-factor 0.25(4) eV b. The abbreviation c.m. means center of mass system.

As it can be seen, there is a visible difference between these results, so, in the future we will generally orient to the total value of the *S*-factor at zero energy which was measured in various works. Furthermore, the new experimental data[33] lead to the value of total *S*(0)=0.216(10) eV b and this means that contributions of *M*1 and *E*1 will change as compared with data.[32] The following parameters of linear approximation from Eq. (30) are given in Ref. 33 $S_0$=0.216(6) eV b and $S_1$=0.0059(4) eV b keV$^{-1}$, that are noticeably differ from the data of work Schimd *et al.*[32]

The other known results for the *S*-factor obtained from the experimental data, without separation to *M*1 and *E*1 parts yield for zero energy 0.165(14) eV b.[41] The previous measurements of the same authors lead to the value 0.121(12) eV b,[42] and for theoretical calculations of work Viviani *et al.*[43] the values $S_s$=0.105 eV b, $S_p$=0.0800-0.0865 eV b are obtained for different models.

One can see that the cited experimental data over the last 10-15 years are very ambiguous. These results make it possible to conclude that, most probably, the value of the total *S*-factor at zero energy is in the range 0.109-0.226 eV b. The average of these experimental measurements equals 0.167(59) eV b, which quite agrees with the value calculated here on the basis of the *E*1 transition only.

The dashed line in Fig. 1 shows the calculation result for the *E*1 transition for the potential of the ground state from Eq. (28). The total *S*-factor is shown in Fig. 1 by the solid line, which demonstrates clearly the small contribution of the *M*1 transition to the $S_s$-factor at the energies above 100 keV and its considerable influence in the energy range of 1-10 keV.

The total *S*-factor dependence on energy in the range of 2.5-50 keV is in complete accordance with the findings of works Schimd et al.[32] and Casella et al.[33] and for the $S_s$-factor of the $M1$ transition at 1 keV we obtained the value of 0.077 eV b, which leads to the value of 0.212(5) eV b for the total $S$-factor and which is in a good agreement with the new measurements data from LUNA project (see Casella et al.[33]). And as it can be seen from Fig. 1, at the energies of 1-3 keV the value of the total $S$-factor is more stable than it was for the $E1$ transition and we consider it to be absolutely reasonable to write the result as 0.212 eV b with the error of 0.005.

If Eq. (30) will be used for the $S$-factor parametrization, then it is possible to describe the solid line in Fig. 1 by the parameters $S_0$=0.1909 eV b and $S_1$=0.006912 eV b keV$^{-1}$ in the energy range of 1-100 keV, with the average $\chi^2$=0.055.

If we use the parametrization of the form

$$S(E_{c.m.}) = S_0 + E_{c.m}S_1 + E^2_{c.m}S_2, \qquad (31)$$

then, the next values were obtained for the parameters: $S_0$=0.1957 eV b, $S_1$=0.006055 eV b keV$^{-1}$ and $S_2$=0.00001179 eV b keV$^{-2}$, with the average $\chi^2$=0.017 in the energy range 1-100 keV. The 10% errors of the calculated $S$-factor values are used for determination of $\chi^2$.

The approximation of the calculation results, by the analytical function of certain type with the performing of $\chi^2$ minimization, is actually done here and further, therefore the $S_0$ and $S(0)$ values are slightly differ, but this difference usually not more than 10%. The quadratic form in Eq. (31) reproduces the behavior of the calculated $S$-factor a bit better, as one can see.

There is another method of the $S$-factor parametrization: when the value $S_0$, determining its behavior at zero energy, is predetermined. In this case, these values are obtained for the parameters of the form Eq. (31): $S_0$=0.2120 eV b, $S_1$=4.5366·10$^{-3}$ eV b keV$^{-1}$ and $S_2$=2.8622·10$^{-5}$ eV b keV$^{-2}$ with the average $\chi^2$=0.124, at the same energy range and for 10% errors.

However, it is necessary to note that we are unable to build the scattering $^2S$-potential uniquely, because of the ambiguities in the results of different phase shift analysis of the p$^2$H scattering. The other variant of the potential with parameters $V_0$=-35.0 MeV and $\alpha$=0.1 fm$^{-2}$,[11,24] which also describes well the $S$ phase shift of scattering, leads at these energies to $S$-factor of the $M1$ process several times lower than in the previous case.

Such a big ambiguity in parameters of the $^2S$-potential of scattering, associated with errors of phase shifts extracted from the experimental data, does not allow us to make certain conclusions about the contribution of the $M1$ process in the radiative proton capture on $^2$H. If the GS potentials are determined by the binding energy, asymptotic constant and charge radius quite uniquely and the potential description of the scattering phase shifts, which are "pure" in accordance with Young schemes, is an additional criteria for the determination of such parameters, then, for the construction of the scattering potential, it is necessary to carry out a more accurate phase shift analysis for the $^2S$-wave and to take into account the spin-orbital separation of the $^2P$ phase shifts at low energies, as it was done for the p$^{12}$C elastic scattering at energies 0.2-1.2 MeV.[44] This will allow us to adjust the potential parameters used in the calculations of the proton capture on $^2$H in the potential cluster model.

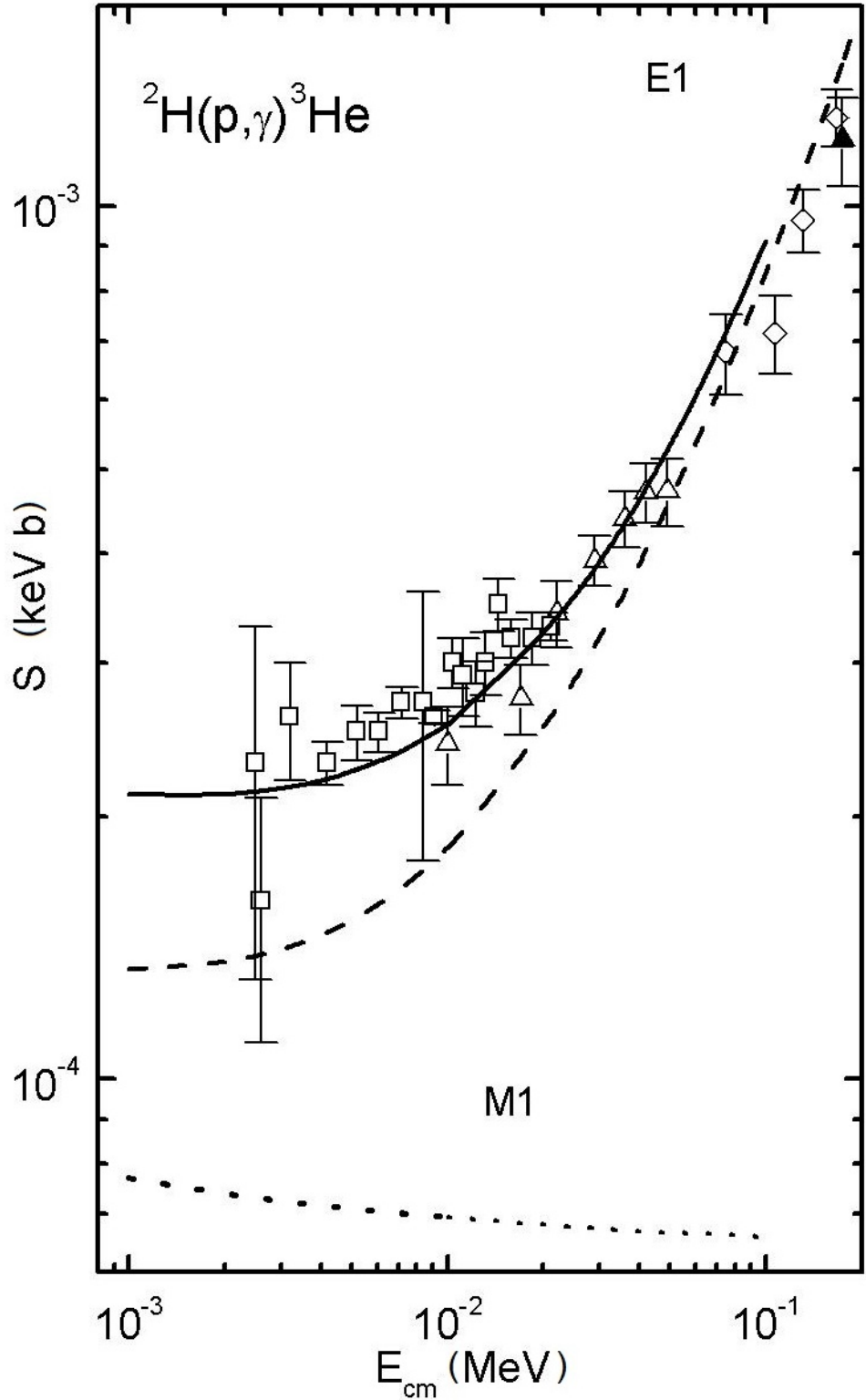

Fig. 1. Astrophysical *S*-factor of the proton capture on $^2$H in the range of 1 keV-0.3 MeV. Lines: calculations with the potentials mentioned in the text. Triangles denote the experimental data from Ref. 30, open rhombs from Ref. 31, open triangles from Ref. 32, open squares from Ref. 33.

Thus, the *S*-factor calculations of the radiative proton capture on $^2$H for the *E*1 transition at the energy range down to 10 keV, which we carried out about 15 years ago,[24] when the experimental data above 150-200 keV were only known, are in a good agreement with the new data from Refs. 31-33 in the energy range 10-150 keV. Moreover, this is true about both the potential from Table 1 and the interaction with parameters from Eq. (28). The results for the two considered potentials at the energies lower than 10 keV practically fall within the error band of work Casella et al.[33] and show that the S-factor tends to remain constant at energies 1-3 keV. In our calculations[24] there was intrinsically predicted the behavior of the S-factor of $^2$H(p, γ)$^3$He in the energy range from 10-20 to 150-200 keV, which value is generally defined by the E1 transition at these energies.

In spite of the uncertainty of the M1 contribution to the process, which results from the errors and ambiguity of $^2$S-phases of scattering, the scattering potential (set forth in Table 1) with mixed Young schemes in the $^2$S-wave allows obtaining a reasonable value for the astrophysical $S_s$-factor of the magnetic transition in the range of low energies. At the same time, the value of the total S-factor is in a good agreement with all known experimental measurements at energies from 2.5 keV to 10 MeV.

As a result, the PCM based on the intercluster potentials adjusted for the elastic scattering phase shifts and GS characteristics, for which the FS structure is determined by the classification of BS according to Young orbital schemes and potential parameters suggested as early as 15 years ago,[24] allows one to describe correctly the astrophysical S-factor for the whole range of energies under consideration.

## 4. The radiative proton capture on $^6$Li

In this section, on the basis of new experimental measurements from Refs. 45-47 and differential cross-sections of the elastic scattering at the energy 500 keV from earlier work, Skill et al.,[48] we have carried out the phase shift analysis and obtained $^{2,4}$S and $^2$P-phase shifts of scattering. The p$^6$Li interaction potentials for L=0 and 1 and without taking into account the spin-orbital separation were constructed according to the found phase shifts, and then the calculations of the astrophysical S-factor at the energy above 10 keV were made.[47]

Although, the $^6$Li(p, γ)$^7$Be radiative capture reaction may be of certain interest for nuclear astrophysics,[2,3,5] it has not been experimentally studied sufficiently well. There are a comparatively small number of measurements of the total cross-sections and calculations of the astrophysical S-factor,[10] and they were performed only in the energy range from 35 keV to 1.2 MeV. Nevertheless, it would be interesting to consider the possibility of description of the S-factor in the frame of the potential cluster model, taking into account the classification of the bound states according to the orbital Young schemes at the astrophysical energy range, where the experimental data exist.

### 4.1. *Potential description of scattering phase shifts*

We have done the p$^6$Li phase shift analysis of the elastic scattering in Ref. 49 and the general pattern of the $^2$S and $^4$S-phase shifts of scattering is shown in Fig. 2. In spite of the large data spread for the $^4$S-phase shifts, the doublet $^2$S-phase shift tends to decrease, but significantly slower than it could be expected from the results of analysis,[50] where the $^2$P-wave was not taken into account.

Errors of the elastic phase shifts of scattering, which are shown in Fig. 2, are due to the ambiguity of the phase shift analysis - it is possible to obtain slightly different values of the phase shifts of scattering with approximately the same value of $\chi^2$. This ambiguity is estimated as $1^0$-$1.5^0$ and is shown for the $S$ and $^2P$-phase shifts in Fig. 2.

Now, let us consider the classification of the orbital states of the p$^6$Li system. The possible orbital Young schemes are listed in Table 3, if the orbital schemes {6} and {42} are used for $^6$Li in the $^2$H$^4$He channel.

Table 3. The classification of the orbital states in p$^6$Li and n$^6$Li systems.

| System | $T$ | $S$ | $\{f\}_T$ | $\{f\}_S$ | $\{f\}_{ST}=\{f\}_S\otimes\{f\}_T$ | $\{f\}_L$ | $L$ | $\{f\}_{AS}$ | $\{f\}_{FS}$ |
|---|---|---|---|---|---|---|---|---|---|
| n$^6$Li p$^6$Li | 1/2 | 1/2 | {43} | {43} | {7} + {61} + {52} + {511} + {43} + {421} + {4111} +{322} + *{3211}* + *{2221}* +{331} | {7} {61} {52} *{43}* *{421}* | 0 1 0,2 1,3 1,2 | – – – {43} {421} | {7} {61} {52} – – |
| | | 3/2 | {43} | {52} | {61} + {52} + {511} + {43} + 2{421} + {331} + {322} + *{3211}* | {7} {61} {52} {43} *{421}* | 0 1 0,2 1,3 1,2 | – – – – {421} | {7} {61} {52} {43} – |

*Note*: $T$, $S$ and $L$ are, respectively, the isospin, spin and orbital moments of particles in the p$^6$Li system; $\{f\}_S$, $\{f\}_T$, $\{f\}_{ST}$ and $\{f\}_L$ are, respectively, the spin, isospin, spin-isospin and possible orbital Young schemes; $\{f\}_{AS}$ and $\{f\}_{FS}$ are the Young schemes of, respectively, allowed and forbidden states. The conjugate schemes $\{f\}_{ST}$ and $\{f\}_L$ are shown in boldface italic.[51]

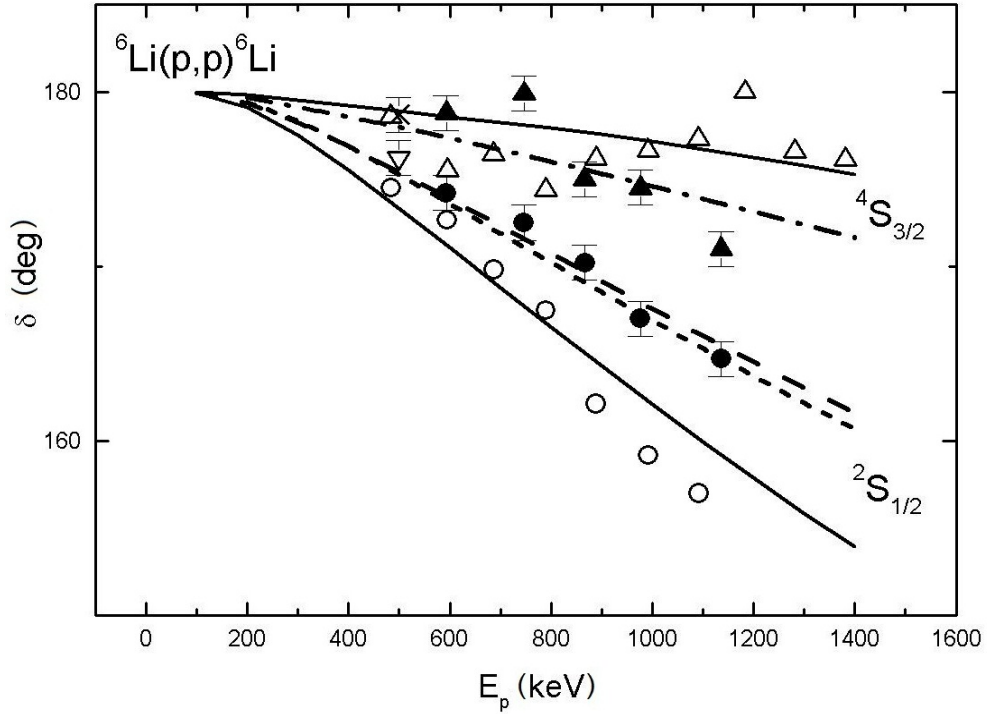

Fig. 2. Doublet and quartet $S$-phase shifts of the elastic p$^6$Li scattering at low energies. Doublet and quartet $S$-phase shifts taking into account $^2P$-wave, when $^4P$-phase shift was taken to be zero. $^2S$ (points) and $^4S$ (triangles) phase shifts[2,3] are obtained on the basis of data.[45,46] The results from Ref. 49 are represented by open triangles and circles, for comparison. Lines show calculation results for different potentials.

As it is seen from Table 3, there are two allowed schemes {43} and {421} in the doublet spin state of the p$^6$Li system and then scattering states are "mixed" according to orbital symmetries. At the same time, one generally consider that for the doublet GS of $^7$Be in the p$^6$Li channel with $J=3/2^-$ и $L=1$ corresponds only one allowed scheme {43}.[6,53]

The considered p$^6$Li system is completely analogous to the p$^2$H channel in $^3$He, which doublet state is also "mixed" according to the Young schemes {3} and {21}. Therefore, the potentials, which are constructed on the base of the phase shifts of the elastic scattering description in the p$^6$Li system can not use for the description of the GS of $^7$Be in the p$^6$Li channel. In this case, the phase shifts of the p$^6$Li elastic scattering as well as p$^2$H system are represented as the half-sum of the "pure" phase shifts in Eq. (21).

$$\delta_L^{\{43\}+\{421\}} = 1/2\delta_L^{\{43\}} + 1/2\delta_L^{\{421\}}. \qquad (32)$$

The "mixed" phase shifts are determined as a result of the phase shift analysis of the experimental data, which are, generally, differential cross-sections of the elastic scattering or excited functions. Then, it is supposed that it is possible to use the phase shifts, of the same symmetry from the quartet channel, as the {421} "pure" phase shifts of the doublet channel. As a result, it is possible to find the {43} pure doublet phase shifts of the p$^6$Li scattering and use these phase shifts for the construction the pure interaction, which ought to correspond to the potential of the bound state of the p$^6$Li system in $^7$Be.[6]

However, it was found that such potential, based on these principles, gives the wrong binding energy of $^7$Be in the p$^6$Li channel, i.e. if we use the methods of receiving the "pure" phase shifts, given in Ref. 6, 55, it is not possible to obtain the "pure" in Young schemes potential of the ground state. It seams that it is due to the absence of the spin-orbital separation and the small probability of clusterization of $^7$Be into the p$^6$Li channel and the other methods will be used for the construction of the GS potential.

Let us consider, at first, the construction of the partial intercluster p$^6$Li interactions of scattering according to the existing phase shifts, we use common Gaussian potential with a point-like Coulomb component, which can be represented as Eq. (14). The following parameters for the description of the results of the phase shift analysis of work Petitjean et al.[50] were obtained:

$$^2S - V_0=-110 \text{ MeV}, \alpha=0.15 \text{ fm}^{-2}, \qquad (33)$$

$$^4S - V_0=-190 \text{ MeV}, \alpha=0.2 \text{ fm}^{-2}. \qquad (34)$$

They include two forbidden bound states which correspond to the Young schemes {52} and {7}.[11] The calculation results of the phase shifts for these potentials are shown in Fig. 2 by the solid lines with the results of the phase shift analysis[50] shown by blank circles and blank triangles.

For the description of our results of the phase shifts of scattering, the potentials with following parameters are preferable:

$$^2S - V_0=-126 \text{ MeV}, \alpha=0.15 \text{ fm}^{-2}, \qquad (35)$$

$$^4S - V_0=-142 \text{ MeV}, \alpha=0.15 \text{ fm}^{-2}. \qquad (36)$$

They also include two forbidden bound states with schemes {52} and {7}. The phase shifts for these potentials are shown in Fig. 2 by the dashed and dot-dashed lines in comparison with the results of our phase shift analysis given by black points and triangles.[49]

Therefore, the "pure", according to orbital symmetries, $^2P_{3/2}$-potential of the ground state of $^7$Be with Young scheme {43} was constructed in such a way that the channel energy, i.e., the binding energy of the ground state of the nucleus with $J=3/2^-$ as p$^6$Li system and its mean square charge radius, were described well. The thus-obtained parameters of the "pure" $^2P_{3/2}^{\{43\}}$-potential can be represented as

$$^2P_{3/2} - V_P=-252.914744 \text{ MeV}, \alpha_P=0.25 \text{ fm}^{-2}. \qquad (37)$$

This potential yields the binding energy of an allowed state (AS) with scheme {43} equal to -5.605800 MeV, while the experimental value is equal to -5.6058 MeV[56] and has one forbidden state corresponding to Young scheme {61}. The mean square charge radius is equal to 2.63 fm which is generally in agreement with the data of Ref 56, and the $C_W$ constant in Eq. (16) is equal to 2.66(1) within the range of 5-13 fm.

For the parameters of $^2P_{1/2}^{\{43\}}$-potential of the first excited state of $^7$Be with $J=1/2^-$ the next values were obtained

$$^2P_{1/2} - V_P=-251.029127 \text{ MeV}, \alpha_P=0.25 \text{ fm}^{-2}. \qquad (38)$$

This potential leads to the binding energy of -5.176700 MeV while its experimental value is equal to -5.1767 MeV[56] and it contains the forbidden state with scheme {61}. Asymptotic constant in Eq. (16) is equal to 2.53(1) within the range of 5-13 fm and the mean square charge radius is equal to 2.64 fm. The absolute accuracy of finding the binding energy in our new computer programs was taken to be at the level of 10$^{-6}$ MeV.[4]

The obtained parameters of potentials of bound states differ slightly from our previous results.[51] This is because we used the exact mass values of particles and more accurate description of the experimental values of the energy levels in these calculations.

It should be noted that, on the basis of the obtained phase shift analysis for the quartet $^4S$-phase shift of scattering that is shown in Fig. 2, it is impossible to construct the unique $^4S$-potential. The results of the phase shift analysis at higher energies are required and it is necessary to take into account the spin-orbital phase shift separation.

Only the $^2S$-interaction is obtained quite uniquely. This interaction with the above $^2P$-potentials of the bound states can be used in future, for example, for the calculations of the astrophysical $S$-factor with the $E1$ transition from the doublet $^2S$-wave of scattering to the ground and first excited doublet bound $^2P$-states of $^7$Be.

The variational method, used for the energy of the ground state, gives the value -5.605797 MeV and hence, the average energy for this potential is -5.6057985(15) MeV, i.e. the accuracy of its determination equals ±1.5 eV. The asymptotic constant at the distances of 5-13 fm turned out to be comparatively stable and is equal to 2.67(2) and the charge radius is in agreement with the calculation results based on the finite-difference method. The expansion parameters of the variational wave function from Eq. (15), for the ground state of $^7$Be in the p$^6$Li channel with potential from Eq. (37), are listed in Ref. 4, and the residual error does not exceed 10$^{-12}$.

The variational method yielded an energy of -5.176697 MeV for the first excited level; therefore, the average energy is equal to -5.1766985(15) MeV, with the same accuracy as it was in the case of the GS. The asymptotic constant at distances of 5-13 fm turned out to be of the level of 2.53(2), the residual error being not more than 10$^{-12}$ and the charge radius being almost the same as for the GS. The parameters of the excited state of the WF of $^7$Be with potential from Eq. (38) are listed in Ref. 4.

## 4.2. *Astrophysical S-factor*

The $E1$ transitions from $^2S$ and $^2D$-states of scattering to the ground $^2P_{3/2}$ and the first excited $^2P_{1/2}$ bound states of $^7Be$ were taken into account when the astrophysical $S$-factor of the radiative proton capture on $^6Li$ was considered. The calculation of the wave function of the $^2D$-wave without spin-orbital separation was made on the basis of the $^2S$-potential but with the orbital moment $L=2$.[57]

When the calculations were made, it turned out that the $^2S$-potential of scattering with the depth of 110 MeV given above and based on the phase shift analysis[50] led to the astrophysical $S$-factor significantly lower than it had to be. At the same time, the doublet $^2S$-potential with the depth of 126 MeV, follows from our results of the phase shift analysis,[49] gives a correct description of the general behavior of the experimental $S$-factor.[58-60] The obtained results are shown in Fig. 3. The results of the transitions from $^2S$ and $^2D$-waves of scattering to the ground state of $^7Be$ are shown by the dashed line, the dotted line is for the transitions to the first excited state and the solid line is the total $S$-factor.

The behavior of the $S(1/2^-)$-factor well reproduces the experimental data (circles in Fig. 3) for the transition to the first excited state of $^7Be$ at low energies. The calculated $S$-factor at 10 keV is equal to $S(3/2^-)=76$ eV b and $S(1/2^-)=38$ eV b while the total value is equal to 114 eV b.

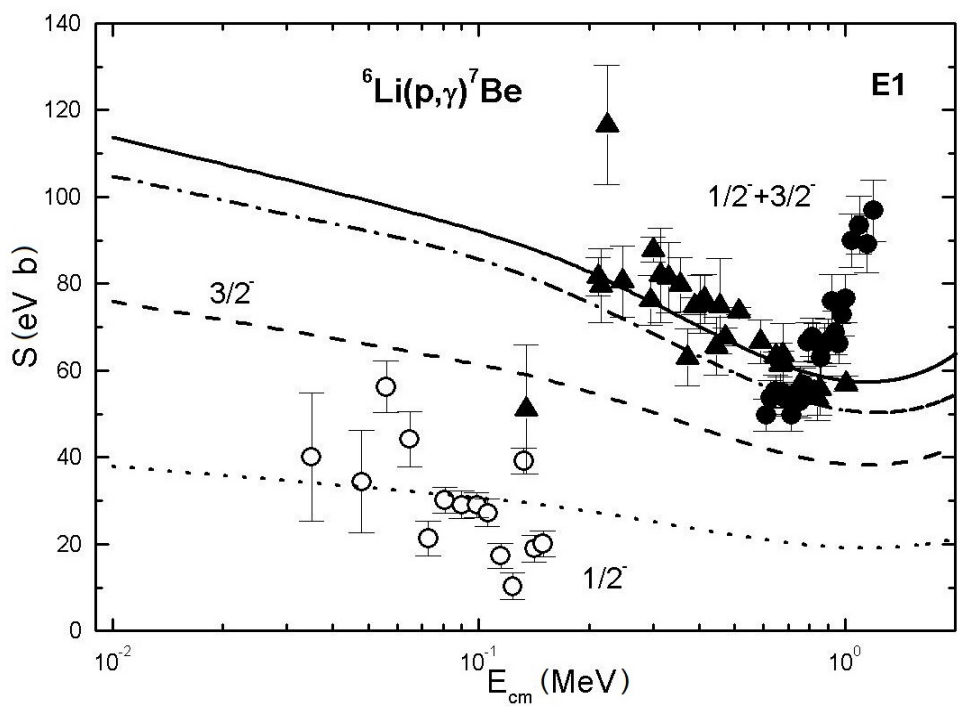

Fig. 3. Astrophysical $S$-factor of the radiative proton capture on $^6Li$. Black points, triangles and circles are the experimental data from Ref. 57 given in Ref. 58, 59. The result for transitions from $^2S$ and $^2D$-waves of scattering to the ground state of $^7Be$ is shown by the dashed line and for transitions to the first excited state - by the dotted line. The solid line shows the total $S$-factor. Dot-dashed line is the result for other variant of the scattering potential.

For comparison of the calculated $S$-factor at zero energy (energy of 10 keV is taken as zero), we will give the known results for the total $S(0)$ factor: 79(18) eV b,[61] 105 eV b (at 10 keV)[60] and 106 eV b.[62] In Ref. 63 for the $S$-factor of transitions to the ground state the value of 39 eV b is given and for the transition to the first excited state the value of 26 eV b, so the total *S*-factor is equal to 65 eV b. As it is seen the difference between these data is comparatively large, and our results are in agreement with them in general.

Moreover, a slight change of the depth of $^2S$ scattering potential, which virtually has no impact on the behavior of calculated phases, rather strongly influences the *S*-factor. For example, for the potential depth equal to 124 MeV, the phases are shown in Fig. 2 by short-dashed lines, and for energy of 10 keV we obtain 105 eV b for the *S*-factor, that is in a good agreement with the latest experimental data.[60,61] The total *S*-factor with this potential is shown in Fig. 3 by the dot-dashed line which lies within the experimental error band at energies below 1 MeV.

The dot-dashed line in Fig. 3, with the average $\chi^2=0.0048$ and for 10% errors, can be parameterized in the energy range 1-200 keV by the expression:

$$S(E_{c.m.}) = S_0 + E^{1/2}_{c.m} S_1 , \qquad (39)$$

with the parameters: $S_0$=109.58 eV b and $S_1$= -2.3992 eV b keV$^{-1/2}$. At that, for 1 keV, the value 106 eV b is used as the experimental value of the *S*-factor. At the second method of the parametrization, considered in the previous section, the next values were obtained: $S_0$=106.00 eV b and $S_1$=-2.0647 eV b keV$^{-1/2}$ with the average $\chi^2$=0.0230.

It should be mentioned that, if we use the potentials without the forbidden states in *S* and *P*-waves or with another number of FS, then the value of the calculated *S*-factor turns out to be from 3 to 100 times lower that the values obtained above. For example, the $^2S$-potential with one forbidden state and parameters -25 MeV and 0.15 fm$^{-2}$, which also gives a good description of the phase shifts of scattering and the potential of the ground state given above, leads to the *S*-factor of about 1 eV b.

Thus, the doublet $^2S$-phase shifts obtained in our phase shift analysis that takes into account the doublet $^2P$-phase shift, lead to the potential which allowed to describe the experimental *S*-factor at energies below 1 MeV, in distinction from the interaction constructed on the basis of the analysis results.[50] The potential cluster model with the above potentials makes it possible to obtain quite reasonable results for the description of the process of the radiative proton capture on $^6$Li in the astrophysical energy region,[41] as it was in the case of lighter nuclei.[64]

## 5. The radiative proton capture on $^7$Li

The radiative capture reaction

$$p + {^7}\text{Li} \rightarrow {^8}\text{Be} + \gamma \qquad (40)$$

at ultralow energies resulting in formation of unstable $^8$Be nucleus which decays into two α-particles may take place along with the weak process

$$^7\text{Be} + e^- \rightarrow {^7}\text{Li} + \gamma + \nu_e , \qquad (41)$$

as one of the final reactions of the proton-proton chain.[5] Therefore, the in-depth study of this reaction, in particular of the form and energy dependence of the astrophysical *S*-factor, is of a certain interest for the nuclear astrophysics.

It is necessary to know the partial potentials of the p$^7$Li interaction in continuous and discrete spectra for calculation of the astrophysical *S*-factor for the radiative proton capture on $^7$Li in the potential cluster model,[6,11] which we usually use for such calculations.[65] We will consider that such potentials should correspond to the classification of cluster states by orbital symmetries[6] as it was assumed in our earlier works.[22,66]

Let us note, that the potentials of scattering processes are constructed on the basis of description of the phase shifts of elastic scattering obtained from the experimental data, while the interactions in bound states are determined from the requirement to reproduce the main characteristics of the bound state of nucleus, if it is mainly due to cluster channel consisting of the initial particles of the considered reaction.

For example, colliding at low energies particles in the $^4$He($^2$H, γ)$^6$Li radiative capture process are formed the $^6$Li nucleus in the ground state and the excess energy is released as a γ quantum. We can consider potentials of one and the same nuclear system of particles, i.e. the $^2$H$^4$He system in continuous and discrete spectra, since there is no restructuring in such reactions. In the last case, it is assumed that the ground state of $^6$Li is very likely caused by the $^2$H$^4$He cluster configuration. Such approach leads to quite reasonable results in the description of the astrophysical S-factors of this and some other radiative capture reactions.[65]

It seems that in this case, $^8$Be can not be examined only as the p$^7$Li cluster system, and most probably it is determined by the $^4$He$^4$He configuration into which it decays. However, it is possible that $^8$Be is in the bound state (p$^7$Li channel), for a while, just after of the radiative proton capture on $^7$Li, and only after this it changes to the state defined by the unbound $^4$He$^4$He system. Such an assumption makes it possible to consider $^8$Be as the p$^7$Li cluster system and use PCM methods, at least at the initial stage of the formation of this nucleus in $^7$Li(p, γ)$^8$Be.[66]

### 5.1. *Classification of the orbital states*

At first, we would like to note that the p$^7$Li system has the $T_z = 0$ isospin projection and it is possible for two values of total isospin $T = 1$ and 0,[67] therefore p$^7$Li channel is mixed by isospin as p$^3$H system,[64] even though, as it will be shown later, both of isospin states ($T = 1,0$), in contrast to p$^3$H system, in the triplet spin state correspond to the allowed Young scheme {431}.[11] The cluster p$^7$Be and n$^7$Li channels with $T_z = \pm 1$ and $T = 1$ are pure by isospin in complete analogy with the p$^3$He and n$^3$H systems.[68,69]

The spin-isospin Young schemes of $^8$Be for the p$^7$Li channel are the product of spin and isospin parts of schemes of the WF. Particularly, under consideration of any of these moments we will have {44} scheme at the ground state of $^8$Be with the moment equals zero, scheme {53} for a certain state with moment equals one and for the state with the moment equals two – {62} symmetry.

The possible Young schemes of the p$^7$Li system turn out to be forbidden if the scheme {7} is used for $^7$Li, because of the rule indicates that there can not be more than four cells in a row,[53,68,69] and they correspond to forbidden states with configurations {8} and {71} and relative motion moments $L = 0$ and 1, which is determined by Elliot rule.[53] The p$^7$Li system, in the triplet spin state, contains forbidden states with the scheme {53} in $P_1$-wave and {44} in $S_1$-wave and allowed state with the configuration {431} at $L = 1$ when the scheme {431} is accepted for $^7$Li.[70]

Thus, the p$^7$Li potentials in the different partial waves should have the forbidden bound state {44} in the $S_1$-wave and forbidden and allowed bound levels in the $P_1$-wave with schemes {53} and {431}, respectively. The considered classification is true for any isospin state of the p$^7$Li system ($T = 0$ or 1) in the triplet spin channel. Allowed symmetries are absent for spin $S = 2$, and all Young schemes, listed above, correspond to the forbidden states.

Possibly, as it was in a previous case for the p$^6$Li system, it is more correctly to consider both allowed schemes {7} and {43} for the bound states of $^7$Li because of the fact that they are present as forbidden and allowed states in the $^3$H$^4$He configuration of this nucleus.[65] Then the level classification will be slightly different, the number of forbidden states will increase, and the extra forbidden state will appear in each partial wave. That kind of more complete scheme of FS and AS states, intrinsically, is a sum of the first and second cases, considered above; and it is listed in Table 4.

Table 4. Classification of the orbital states in the p$^7$Li (n$^7$Be) systems with the isospin $T$=0,1.[70]

| System | T | S | {f}$_T$ | {f}$_S$ | {f}$_{ST}$ = {f}$_S \otimes$ {f}$_T$ | {f}$_L$ | L | {f}$_{AS}$ | {f}$_{FS}$ |
|---|---|---|---|---|---|---|---|---|---|
| p$^7$Li n$^7$Be | 0 | 1 | {44} | {53} | {71} + {611} + {53} + {521} + {431} + {4211} + {332} + *{3221}* | {8} {71} {53} {44} *{431}* | 0 1 1,3 0,2,4 1,2,3 | - - - - {431} | {8} {71} {53} {44} - |
| | | 2 | {44} | {62} | {62} + {521} + {44} + {431} + {422} + {3311} | {8} {71} {53} {44} {431} | 0 1 1,3 0,2,4 1,2,3 | - - - - - | {8} {71} {53} {44} {431} |
| p$^7$Be n$^7$Li p$^7$Li n$^7$Be | 1 | 1 | {53} | {53} | {8} + 2{62} + {71} + {611} + { 53} + {44} + 2{521} + {5111} + {44} + {332} + 2{431} + 2{422} + {4211} + {3311} + *{3221}* | {8} {71} {53} {44} *{431}* | 0 1 1,3 0,2,4 1,2,3 | - - - - {431} | {8} {71} {53} {44} - |
| | | 2 | {53} | {62} | {71} + {62} + {611} + 2{53} + 2{521} + 2{431} + {422} + {4211} + {332} | {8} {71} {53} {44} {431} | 0 1 1,3 0,2,4 1,2,3 | - - - - - | {8} {71} {53} {44} {431} |

*Note*: T, S and L are, respectively, the isospin, spin and orbital moments of particles; {f}$_S$, {f}$_T$, {f}$_{ST}$ and {f}$_L$ are the spin, isospin, spin-isospin and possible orbital Young schemes, respectively; {f}$_{AS}$ and {f}$_{FS}$ are the Young schemes of the allowed and forbidden states, respectively. The conjugate schemes are printed in boldface italic.

## 5.2. *Potential description of scattering phase shifts*

The p$^7$Li elastic scattering is represented as a half-sum of the isospin pure phase shifts[66] because of the isospin mixing of phase shifts, which is in complete analogy with the p$^3$H system considered in Refs. 64, 68, 69. The phase shifts with $T = 1,0$ mixed by isospin are usually determined as a result of the phase shift analysis of experimental data of the elastic scattering differential cross-sections or excitation functions. The pure phase shifts with isospin $T = 1$ are determined from the phase shift analysis of the p$^7$Be or n$^7$Li elastic scattering. As a result, it is possible to find the pure p$^7$Li phase shifts of scattering with $T = 0$ and construct the interaction model using these results, which have to correspond to the potential of the bound state of the p$^7$Li system in $^8$Be.[71] Just the same method of the phase shift separation was used for the p$^3$H system[68,69] and its absolute validity was shown.[11]

However, we failed to find experimental data of differential cross-sections or phase shifts of the p$^7$Be or n$^7$Li elastic scattering at astrophysical energies, so here we will consider only isospin-mixed potentials of the elastic scattering processes in the p$^7$Li system and pure potentials of the bound state with $T = 0$, which are constructed on the basis of description of the BS characteristics and are chosen in the Gaussian form with the point-like Coulomb term (see Eq. (14)).

The phase shifts of the p$^7$Li elastic scattering obtained from the phase shift analysis of the experimental data of excitation functions[72] taking into account spin-orbital separation at energies below 2.5 MeV are given in work Brown *et al.*[73] Later, we will use these phases of the p$^7$Li elastic scattering, which are equal to zero for the $S_1$-wave at energies below 600-700 keV, for the construction of the intercluster potential in $S_1$- and $P_1$-waves. Since, later we will consider the low and astrophysical energy range only, so we will come to nothing more than the energy interval from 0 keV to 700 keV. Practically zero phase shift at these energies is obtained with the potential of the form, given in Eq. (14) and with parameters:

$$V_0 = -147.0 \text{ MeV and } \alpha = 0.15 \text{ fm}^{-2}. \tag{42}$$

Such potential contains two FS as it follows from the state classification given above. Of course, the $S_1$-phase shift at about zero one can obtain from the other variants of potential parameters with two FS. In this regard, it is not possible to fix its parameters unambiguously, and the other combinations of $V_0$ and $\alpha$ are possible. However, this potential, as the potential given above, should has comparatively large width, which gives small phase shift variation, when the energy changes in the range from 0 to 700 keV.

There is the over-threshold level in the $P_1$-wave with the energy 17.640 MeV and $J^PT = 1^+1$ or 0.441 MeV in laboratory system (l.s.) which is above the threshold of the p$^7$Li cluster channel in $^8$Be, with the binding energy of this channel -17.2551 MeV.[56] The 0.441 MeV level has the very small width, that for the $^7$Li(p, $\gamma$)$^8$Be reaction and for the p$^7$Li elastic scattering, equals only 12.2(5) keV.[56] Such narrow level leads to the sharp rise of the $P_1$-phase shift of elastic scattering, which should be mixed by spin states $^5P_1$ and $^3P_1$ for the total moment $J = 1$.[56] The phase shift, which is shown by points in Fig. 4,[73] is described by the Gaussian potential from Eq. (14) with parameters:

$$V_0 = -5862.43 \text{ MeV and } \alpha = 3.5 \text{ fm}^{-2}. \tag{43}$$

This potential, mixed in isospin $T = 0$ and 1, has two FS and the calculation results of the $P_1$-phase shift of the elastic scattering are shown in Fig. 4 by solid line.

The potential parameters which describe the $P_1$-phase shift are fixed quite unambiguously at the interval of sharp increase obtained from the experimental data and the potential itself should have very small width.

Since, later we will consider astrophysical $S$-factor only at the energies from 0 to 700 keV, it can be deemed that both of potentials, obtained above, give a good description for the results of the phase shift analysis for two considered partial waves in this energy range.

The following parameters of the potential of the bound $^3P_0$-state of the p$^7$Li system corresponding to the ground state of $^8$Be in the examined cluster channel are obtained:

$$V_0 = -433.937674 \text{ MeV and } \alpha = 0.2 \text{ fm}^{-2}. \quad (44)$$

The binding energy -17.255100 MeV with the accuracy of $10^{-6}$ MeV, the mean square charge radius is equal to 2.5 fm and the asymptotic constant $C_W$=12.4(1), calculated for Whittaker functions from Eq. (16), were obtained with this potential. The error of the constant is estimated by its averaging in the range 6-10 fm where the asymptotic constant is practically stable. In addition to the allowed BS corresponding to the ground state of $^8$Be, such $P$-potential has two FS in total correspondence with the classification of the orbital cluster states.

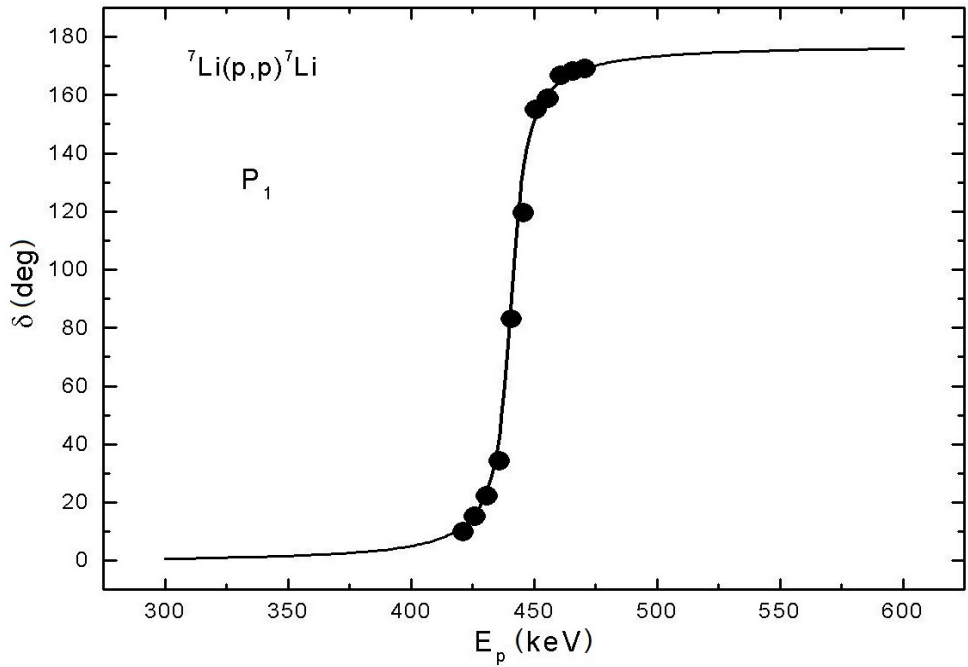

Fig. 4. $^5P_1$-phase mixed with $^3P_1$-phase of the p$^7$Li elastic scattering at low energies. Points - phase shifts obtained from the experimental data of Brown et al.[73] Line - calculations with the Gaussian potential based on parameters given in the text.

It seems that the mean square charge radius of $^8$Be in the p$^7$Li cluster channel should not differ a lot from the radius of $^7$Li, which equals 2.35(10) fm,[56] since the nucleus is in a strongly bound (~ -17 MeV), i.e. compact state. Moreover, at such binding energy of the nucleus of $^7$Li itself can be in deformed, compressed form as it is for deuteron nucleus in $^3$He.[71] Therefore, the value of the mean square charge radius for the p$^7$Li channel in the GS of $^8$Be obtained above has a quite reasonable value.

The variational method, with the expansion of cluster wave function of the p$^7$Li system in nonorthogonal Gaussian basis,[20] is used for the additional control of the accuracy of binding energy calculations, and the energy -17.255098 MeV with $N$=10 order of matrix were obtained for this potential, which is differ from the given above finite-difference value by 2 eV only. Residuals[20] are of the order of $10^{-11}$, asymptotic constant at the range 5-10 fm equals 12.3(2), the charge radius does not differ from the previous results. Expansion parameters of the obtained variational GS radial wave function of $^8$Be in the p$^7$Li cluster channel are listed in Ref. 4.

As it was told before, the variational energy decreases as the dimension of the basis increases and gives the upper limit of the true binding energy, but the finite-difference energy increases as the size of steps decreases and the number of steps increases,[20] therefore it is possible to use the average value -17.255099(1) MeV for the real binding energy in this potential. Thus, the calculation error of the binding energy of $^8$Be in the p$^7$Li cluster channel, using two different methods, is about ±1 eV.

### 5.3. *Astrophysical S-factor*

While considering electromagnetic transitions for the *S*-factor calculations, we will take into account the *E*1 process from the $^3S_1$-wave of scattering to the ground bound state of $^8$Be in the p$^7$Li cluster channel with $J^PT = 0^+0$ and the *M*1 transition from the $P_1$-wave of scattering (see Fig. 4), also to the GS of the nucleus. Cross-sections of the *E*1 transition from the $^3D_1$-wave of scattering (with potential for the $^3S_1$-wave at $L$=2) to the GS of $^8$Be are by 2-4 orders lower than from the $^3S_1$-wave transition at the energy range 0-700 keV. Further on, we will consider only the *S*-factor for the transition to the ground state of $^8$Be, i.e. the $^7$Li($p,\gamma_0$)$^8$Be reaction. One of the last experimental measurements of the *S*-factor of this reaction in the energy range from 100 keV to 1.5 MeV was done in work Zahnow *et al.*[74]

The expressions, given above in the first chapter, are used for the *S*-factor calculations. Values: $\mu_p$=2.792847 and $\mu(^7$Li$)$=3.256427 are accepted for magnetic moment of proton and $^7$Li. The calculation results for the *S*-factor, with the given above potentials at the energy range 5-800 keV (l.s.), are shown in Fig. 5. The *E*1 transition is shown by the dashed line, dotted line - *M*1 process, solid line - the sum of these processes. The *M*1 transition in the considered reaction, like the *E*1 transition in the p$^3$H system,[68,69] goes with the isospin change equals $\Delta T$=1, since the ground state of $^8$Be has $T$=0 and the resonance isospin in the $P_1$-wave of scattering equals 1.

The value of 0.50 keV b was obtained for the astrophysical *S*-factor at 5 keV (c.m.) for the transition to the GS of $^8$Be, where the *E*1 process gives the value of 0.48 keV b, which is in a good agreement with the data from Zahnow *et al.*[74] The calculated and experimental *S*-factor values, at the energy range 5-300 keV (l.s.), are given in Table 5. As it can be seen from Fig. 5 and Table 5, the value of the theoretical *S*-factor at the energy range 30-200 keV is almost constant and approximately equal to 0.41-0.43 keV b, which agrees, practically within the experimental errors, with data of Ref. 74 for the energy range 100-200 keV.

Let us compare some extrapolation results of different experimental data to zero energy. The value 0.25(5) keV b was obtained in Cecil *et al.*[75] and the value 0.40(3) keV b in Godwin *et al.*[76] on the basis of data from Ref. 74. Then in Spraker *et al.*[77] on the basis of new measurements of the total cross-sections of $^7$Li(p,$\gamma_0$)$^8$Be at the energy range 40-100 keV the value 0.50(7) keV b was suggested, which is in a good agreement with the obtained above value at the energy 5 keV.

Table 5. Calculated astrophysical *S*-factor of the radiative proton capture on $^7$Li at low energies and its comparison with the experimental data.[74]

| $E_{lab.}$ (keV) | $S_{exp.}$ (keV b)[74] | $S_{E1}$ (keV b) | $S_{M1}$ (keV b) | $S_{E1+M1}$ (keV b) |
|---|---|---|---|---|
| 5.7 | --- | 0.48 | 0.02 | 0.50 |
| 29.7 | --- | 0.41 | 0.02 | 0.43 |
| 60.6 | --- | 0.39 | 0.02 | 0.41 |
| 98.3 | 0.41(3) | 0.39 | 0.03 | 0.42 |
| 198.3 | 0.40(2) | 0.37 | 0.06 | 0.43 |
| 298.6 | 0.49(2) | 0.36 | 0.16 | 0.52 |

One has to use the quadratic approximation from Eq. (31), in order to correctly describe the calculated *S*-factor, shown in Fig. 5 by the solid line. In this case, the next values were obtained for the parameters: $S_0$=0.47218 keV b, $S_1$= -1.1996·10$^{-3}$ keV b keV$^{-1}$ and $S_2$=5.3407·10$^{-6}$ keV b keV$^{-2}$, with the average $\chi^2$=0.072 in the energy range 1-200 keV. The 10% errors of the calculated *S*-factor values are used for determination of $\chi^2$ and the *S*-factor's value equals 0.5 keV b, obtained at zero energy, is used at energy of 1 keV.

At the second method of parametrization, considered in the section 3, we obtain: $S_0$=0.5 keV b, $S_1$= -1.7973·10$^{-3}$ keV b keV$^{-1}$ and $S_2$=8.0259·10$^{-6}$ keV b keV$^{-2}$ with the average $\chi^2$=0.118.

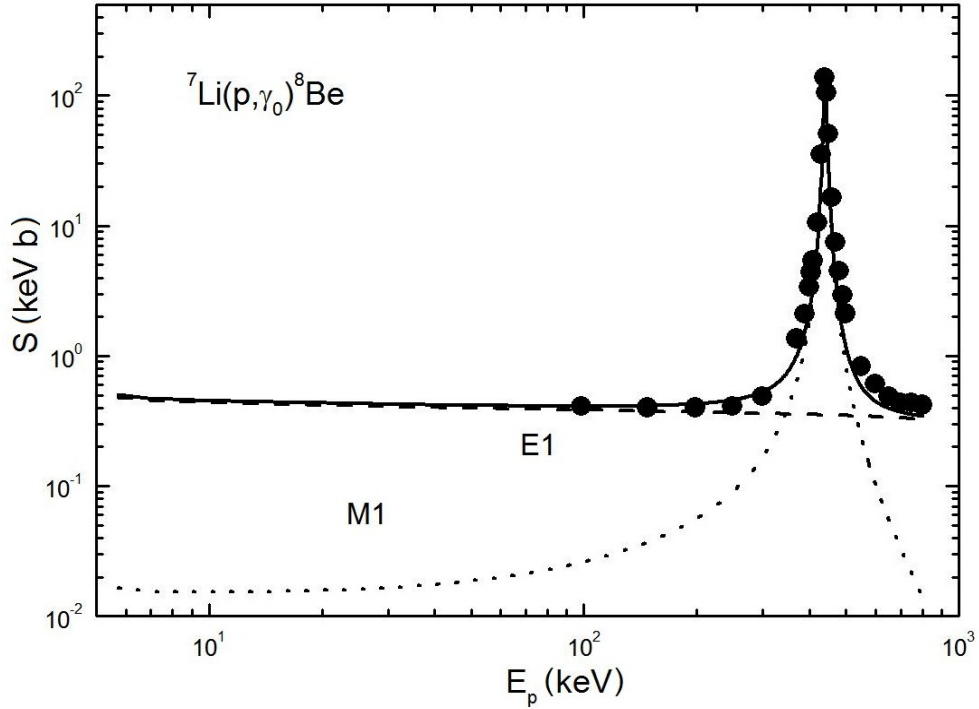

Fig. 5. Astrophysical *S*-factor of the radiative proton capture on $^7$Li. Points: the experimental data from work Zahnow, *et al.*[74] Lines: the calculation results for different electromagnetic transitions with the potentials mentioned in the text.

It is interesting to draw attention at the chronology of different works for obtaining of the astrophysical $S$-factor of the $^7$Li(p,$\gamma_0$)$^8$Be reaction. It was believed in 1992 that its value equals 0.25(5) keV b,[75] the value 0.40(3) keV b[76] was obtained in 1997 on the basis of measurements[74] that were done in 1995 and the measurements in 1999 at lower energies have resulted 0.50(7) keV b.[77] This chronology demonstrates well the constant increase of the obtained values for the astrophysical $S$-factor of the $^7$Li(p,$\gamma_0$)$^8$Be reaction (two fold increase), while decreasing the energy of experimental measurements.

Thereby, the $E$1 and $M$1 transitions from the $S_1$ and $P_1$-wave of scattering to the ground bound state in the p$^7$Li channel of $^8$Be were considered in the potential cluster model. It is possible to completely describe modern experimental data for the astrophysical $S$-factor at the energies up to 800 keV taking into account certain assumptions concerning the channel restructuring in $^8$Be and to obtain its value for zero (5 keV) energy, which is in a good agreement with the latest experimental measurements.

## 6. The radiative proton capture on $^{12}$C

In this section we will consider the p$^{12}$C system and the process of radiative proton capture on $^{12}$C at astrophysical energies. The new measurement of differential cross-sections of the p$^{12}$C elastic scattering, at energies from 200 keV up to 1.1 MeV (c.m.) within the range of $10^0$-$170^0$ with 10% errors, was carried out in Ref. 78,79. Further, the standard phase shift analysis was done and the potential of $S_{1/2}$-state of p$^{12}$C system was reconstructed in this paper on the basis of these measurements,[44] and then the astrophysical $S$-factor at the energies down to 20 keV was considered in the frame of the potential cluster model.[80]

Let us proceed to the immediate describing of the obtained results we would like to note that this process is the first thermonuclear reaction of the CNO-cycle, which took place at a later stage of the stellar evolution, when the partial hydrogen burning occurred. As the hydrogen is burned, the core of the star starts compressing, which results in the increase in pressure and temperature in the star and along with the proton-proton chain the next chain triggers of thermonuclear processes, called CNO-cycle.[2,3,5] The radiative proton capture on $^{12}$C process is a part of the CNO thermonuclear cycle at low energies and gives a considerable contribution to energy efficiency of thermonuclear reactions.[1-3]

### 6.1. *Potentials of the p$^{12}$C interaction*

As it was told before, on the basis of data,[78,79] we have done the phase shift analysis of the p$^{12}$C elastic scattering[44] and the general view of the $S_{1/2}$-phases are shown in Fig. 6, where black points - results of the phase shift analysis for the $S$-phase shift taking into account the $S$-wave only; open squares - results of the phase shift analysis for the $S$-phase shift taking into account $S$ and $P$-waves;[44] dashed line - result of work Jackson *et al.*;[81] solid line - result calculated with potential from Eq. (45). The scattering phase shifts, obtained in such a way, are used further for the construction of intercluster potentials and for calculations of the astrophysical $S$-factor. The existing experimental data of the astrophysical S-factor[10] indicates the presence of the narrow resonance with the width of about 32 keV at the energy 0.422 MeV (c.m.), which leads to the increase in the S-factor by two-three order.

It is interesting to clear the possibility to describe the resonance S-factor on the basis of the PCM with FS and with the classification of orbital states according to Young schemes. The phase shift analysis of the new experimental data[78] of differential cross-sections of the p$^{12}$C elastic scattering at astrophysical energies,[44] which we have shown above, allows one to construct the potentials of the p$^{12}$C interaction for the phase shift analysis of the elastic scattering.

Let us examine the classification of orbital states according to Young schemes in the p$^{12}$C system for the purposes of construction of the interaction potential. The feasible orbital Young schemes in the $N=n_1+n_2$ particle system can be defined in the following way $\{1\}\times\{444\}=\{544\}$ and $\{4441\}$.[52,53] The first of these schemes is consistent with the orbital moment $L=0$ only and is forbidden, because s-shell cannot contain more than four nucleons.[6] The second scheme is allowed with orbital moments 1 and 3,[52,53] the first of which corresponds to the ground bound state of $^{13}$N with $J=1/2^-$. Therefore, the forbidden bound state must be in the potential of the $^2S_{1/2}$-wave and the $^2P$-wave should has only one allowed state at the energy of -1.9435 MeV.[82]

For the calculations of photonuclear processes, the nuclear part of the p$^{12}$C inter-cluster interaction is represented as Eq. (14) with the point-like Coulomb component. The potential of $^2S_{1/2}$-wave is constructed so as to describe correctly the corresponding partial phase shift of the elastic scattering, which has a well defined resonance at 0.457 MeV (l.s.).

Using the results of the phase shift analysis,[44] the $^2S_{1/2}$-potential of the p$^{12}$C interaction, with FS at the energy $E_{FS}=-25.5$ MeV, was obtained together with parameters:

$$V_S = -102.05 \text{ MeV}, \alpha_S = 0.195 \text{ fm}^{-2}, E_{FS} = -12.8 \text{ MeV}. \qquad (45)$$

The calculation results of the $^2S_{1/2}$-phase shift with this potential are shown in Fig. 6 by solid line.

The potential of the bound $^2P_{1/2}$-state has to reproduce correctly the binding energy of $^{13}$N in the p$^{12}$C channel -1.9435 MeV[82] and reasonably describe the mean square charge radius, which probably does not differ significantly from the radius of $^{14}$N, which is equal to 2.560(11) fm.[82] As a result, the following parameters were obtained:

$$V_{GS} = -144.492278 \text{ MeV}, \alpha_{GS} = 0.425 \text{ fm}^{-2}. \qquad (46)$$

The potential gives the binding energy equals -1.943500 MeV and the mean square charge radius $R_{ch}=2.47$ fm. We use the following values for the radii of proton and $^{12}$C: 0.8768(69) fm and 2.472(15) fm.[13,83] The asymptotic constant $C_W$ with Whittaker asymptotics from Eq. (16) was calculated for controlling behavior of WF of the BS at long distances; its value in the range of 5-20 fm equals 1.36(1).

The variational method, using for an additional control of the binding energy calculations, allows to obtain the binding energy of -1.943499 MeV with the residual error being not more than 6 10$^{-14}$ for the first variant of BS potential given in Eq. (46), by using an independent variation of the parameters and the grid having dimension 10. The asymptotic constant $C_W$ of the variational WF, at distances of 5-17 fm, is at the level of 1.36(2). Its variational parameters are listed in Ref. 4. The charge radius does not differ from the value obtained in the FDM calculations.

For the real binding energy in this potential it is possible to use the value -1.9434995(5) MeV, i.e. the calculation error of finding binding energy is on the level of ±0.5 eV, because the variational energy decreases as the dimension of the basis increases and gives the upper limit of the true binding energy, but the finite-difference energy increases as the size of steps decreases and the number of steps increases.

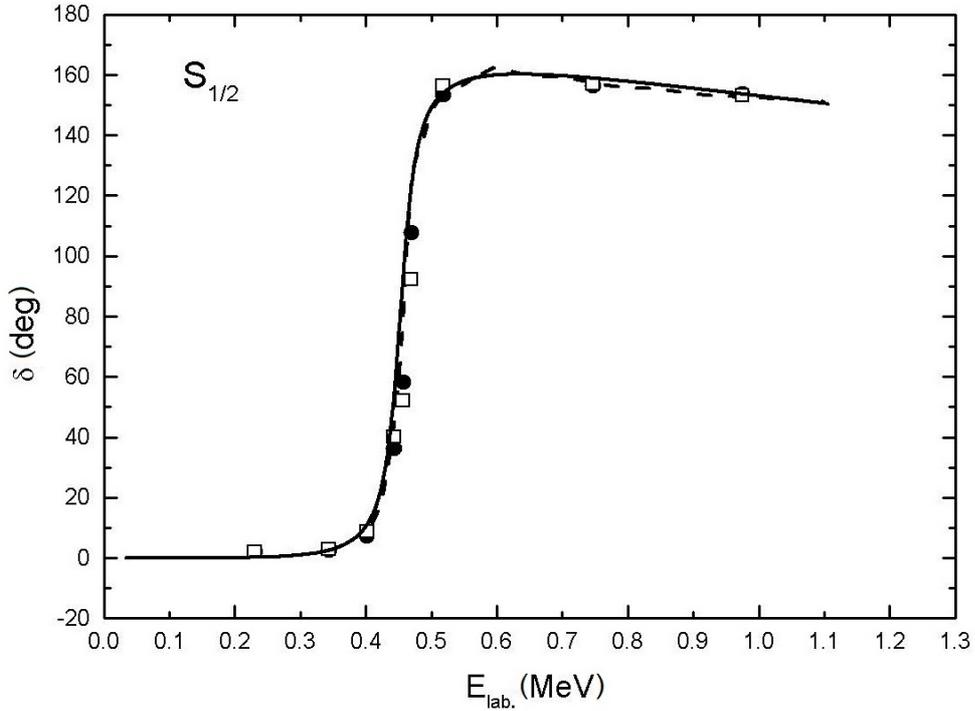

Fig. 6. $^2S$-phase shift of the p$^{12}$C elastic scattering at low energies. Black points - results of the phase shift analysis for the $S$-phase shift taking into account the $S$-wave only; open squares - results of the phase shift analysis for the $S$-phase shift taking into account $S$ and $P$-waves;[44] dashed line - results of Ref. 81; solid lines – results, calculated with the potential from Eq. (45).

### 6.2. *Astrophysical S-factor*

The $E1(L)$ transition resulting from the orbital part of electric operator[11] is taken into account in present calculations of the radiative proton capture on $^{12}$C. The cross-sections of $E2(L)$ and $MJ(L)$ processes and the cross-sections depending on the spin part $EJ(S)$, $M2(S)$ turned out to be a few orders less. The electrical $E1(L)$ transition in $^{12}$C(p, γ)$^{13}$N is possible between the doublet $^2S_{1/2}$ and $^2D_{3/2}$-states of scattering and the ground bound $^2P_{1/2}$-state of $^{13}$N in the p$^{12}$C channel.

It should be noted that, in all calculations, the cross-section of the $E1$ electrical process, caused by the transition from the doublet $^2D_{3/2}$-state of scattering to the ground bound $^2P_{1/2}$-state of $^{13}$N, is lesser than the cross-section of the transition from $^2S_{1/2}$-state of scattering by 4-5 digits. Thus, the main contribution to the calculated $S$-factor of the $^{12}$C(p, γ)$^{13}$N process is made by the $E1$ transition from the $^2S$-wave of scattering to the ground state of $^{13}$N. The mass of proton was taken to be 1 in all calculations for the p$^{12}$C system.

The calculation results of the S-factor of the radiative proton capture on $^{12}$C with the above mentioned potentials for $^2P_{1/2}$ and $^2S_{1/2}$-waves, at energies from 20 keV to 1.0 MeV, are shown together with experimental data from Refs. 10, 84 in Fig. 7 by the solid line. The value 1.52 keV b of the S-factor is obtained at the energy of 25 keV and the extrapolation of the experimental S-factor to 25 keV gives: 1.45(20) keV b and $1.54^{+15}_{-10}$ keV b.[82] Though, in the range of 20-30 keV, the S-factor value is practically constant and one can consider it as the S-factor value at zero energy with an error of about 0.02-0.03 keV b.

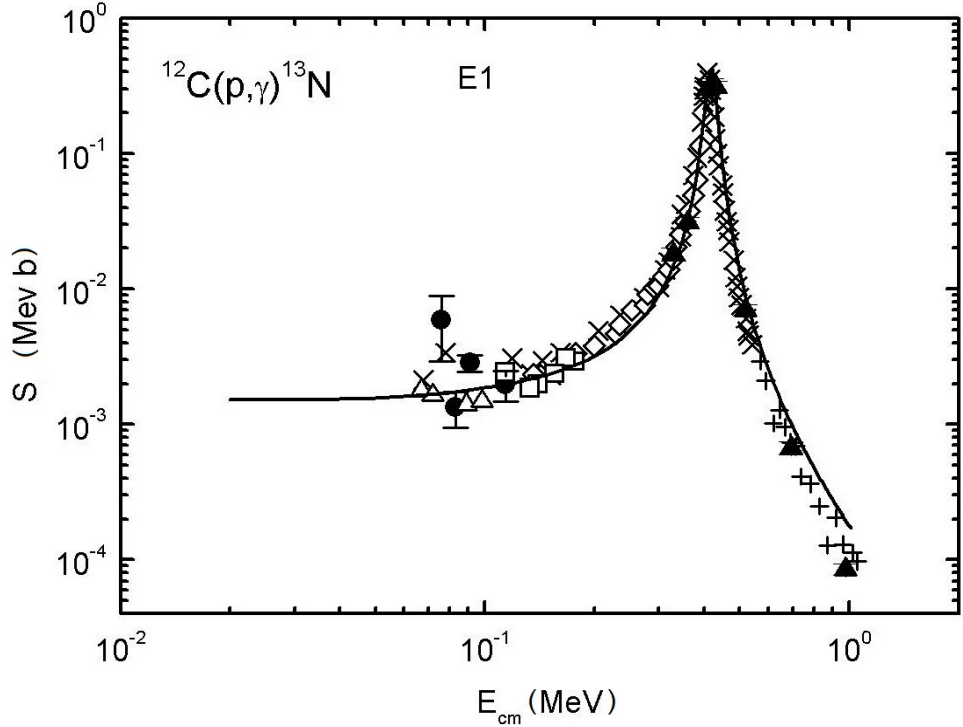

Fig. 7. Astrophysical S-factor of the radiative proton capture on $^{12}$C at low energies. The experimental data, specified as ×, •, □, +, ◊ and Δ, are taken from review,[10] triangles are from Burtebaev et al.[84]. Line: calculations with potentials from Eqs. (45,46).

Thereby, the given above potentials, with FS for the $^2S_{1/2}$-wave and with the bound state without FS that gives the correct binding energy, lead to the joint description of the resonance in the S-factor and the resonance in the $^2S_{1/2}$-phase of scattering.

If Eq. (30) is used for parametrization of the S-factors, then the solid line in Fig. 7 can be described, at the energy range 10-100 keV with the average $\chi^2=0.047$, by the parameters: $S_0=1.4258$ keV b and $S_1=0.003738$ keV b keV$^{-1}$. At that, the S-factor value obtained at 20 keV was used for 10 keV.

The quadratic form in Eq. (31) should to use at the interval 10-200 keV. The next values were obtained for the parameters: $S_0=1.4809$ keV b, $S_1=-1.4894 \cdot 10^{-4}$ keV b keV$^{-1}$ and $S_2=4.0986 \cdot 10^{-5}$ keV b keV$^{-2}$, with the average $\chi^2=0.018$. The 10% errors of the calculated S-factor values are used for determination of $\chi^2$, as before.

At the second method of parametrization, considered in the section 3, we obtain: $S_0=1.52$ keV b, $S_1=-1.0154 \cdot 10^{-3}$ keV b keV$^{-1}$ and $S_2=4.4833 \cdot 10^{-5}$ keV b keV$^{-2}$ with the average $\chi^2=0.0088$.

At the same time, if we use the potentials of the $^2S_{1/2}$-wave with small depth and without forbidden states, for example, with parameters:

$$V_S = -15.87 \text{ MeV}, \quad \alpha_S = 0.1 \text{ fm}^{-2}, \qquad (47)$$

$$V_S = -18.95 \text{ MeV}, \quad \alpha_S = 0.125 \text{ fm}^{-2}, \qquad (48)$$

$$V_S = -21.91 \text{ MeV}, \quad \alpha_S = 0.15 \text{ fm}^{-2}, \qquad (49)$$

then we can't obtain the correct description of the maximum of the *S*-factor of radiative capture reaction. It is impossible to describe the absolute value of the *S*-factor, which, for all variants of the scattering potentials from Eq. (48) and the BS potentials, is in 2-3 times higher than the experimental maximum. At the same time, for all given depthless potentials of the form given in Eq. (48), the resonance behavior of the $^2S_{1/2}$-phase shift of scattering is well described. As the width of the $^2S_{1/2}$-potential decreases, i.e. the $\alpha$ value increases, the value of the *S*-factor maximum grows up, e.g. for the last variant of the $^2S_{1/2}$-scattering potential its value is approximately three times as much as the experimental value.[80]

Thereby, it is possible to describe the astrophysical *S*-factor and the $^2S_{1/2}$-phase shift of scattering in the resonance energy range 0.457 MeV (l.s.) on the basis of the PCM and the deep $^2S_{1/2}$-potential with FS, and to receive the reasonable values for the charge radius and asymptotic constant. The depthless potentials of scattering do not lead to the joint description of the *S*-factor and the $^2S_{1/2}$-phase shift of scattering at any considered combinations of p$^{12}$C interactions.[80]

## 7. The radiative proton capture on $^{13}$C

In this section, we will continue the study of the astrophysical *S*-factors for thermonuclear reactions and will dwell on the $^{13}$C(p, $\gamma$)$^{14}$N radiative capture process at astrophysical energies. The radiative proton capture on $^{13}$C is one of the main reactions of the CNO thermonuclear cycle at low energies and gives a considerable contribution to energy efficiency of thermonuclear reactions.[1-3]

Let us consider the classification of the orbital states of the p$^{13}$C system according to the Young schemes. It was shown earlier that the Young scheme {4441} corresponds to the ground bound state of $^{13}$N so as for $^{13}$C.[17,54,55,80] Let us remind that the possible orbital Young schemes in the $N = n_1 + n_2$ system of particles can be characterized as the direct outer product of the orbital schemes of each subsystem, and, for the p$^{13}$C system within of 1*p*-shell, it yields {1} × {4441} → {5441} + {4442}.[52,53] The first of the obtained scheme is compatible with the orbital moment $L = 1$ and is forbidden, so far as it could not be five nucleons in the *s*-shell, and the second scheme is allowed and is compatible with the orbital moments 0 and 2.[52,53] Thus, in the $^3S_1$-potential there is the only one allowed state and the $^3P$-wave has the forbidden state and allowed one at the energy -7.55063 MeV.[82] However, the above-obtained result should be considered only as qualitative estimation of possible orbital symmetries in the GS of $^{14}$N for the p$^{13}$C channel.

### 7.1. *Phase shift analysis*

While the consideration of the astrophysical *S*-factor of the thermonuclear $^{13}$C(p, $\gamma$)$^{14}$N radiative capture reaction, we have done the phase shift analysis of the p$^{13}$C elastic scattering at the energies from 250 to 800 keV on the basis of the experimental data obtained in Refs. 85, 86. The methods for determination of the elastic scattering cross-section, which are needed for the carrying out of the phase shift analysis, are described in Refs. 14, 87.

One can see that, as the result of the carried out analysis, the singlet $^1S_0$ phase shift is close to zero (within 1°–3°). Fig. 8 shows the form of the triplet $^3S_1$ phase shift. The triplet $^3S_1$ phase shift has the pronounced resonance corresponding to the level $J^PT = 1^-1$ of $^{14}$N in the p$^{13}$C channel at the energy 0.55 MeV (l.s.).[82] The width of this resonance has the value 23(1) keV[82] what less than in case of p$^{12}$C scattering,[44] and we need for its description the narrow potential without the FS what might lead to the width parameter of the order β = 2÷3 fm$^{-2}$.

The nuclear part of the intercluster potential of the p$^{13}$C interaction is represented as usual: in the Gaussian form[88] (see Eq. (14)) with a point-like Coulomb term, given above. The potential for $^3S_1$ wave was constructed so as to describe correctly the resonance partial phase shift of the elastic scattering (see Fig. 8). Two variants for $^3S_1$ potential of the p$^{13}$C interaction, without FS were obtained using the results of our phase shift analysis:

$$V_S = 265.40 \text{ MeV, and } \alpha_S = 3.0 \text{ fm}^{-2}, \tag{50}$$

$$V_S = 186.07 \text{ MeV, and } \alpha_S = 2.0 \text{ fm}^{-2}. \tag{51}$$

The calculation results of $^3S_1$ phase shift with such potentials practically coincide and are shown in Fig. 8 by the solid Eq. (50) and dashed Eq. (51) lines.

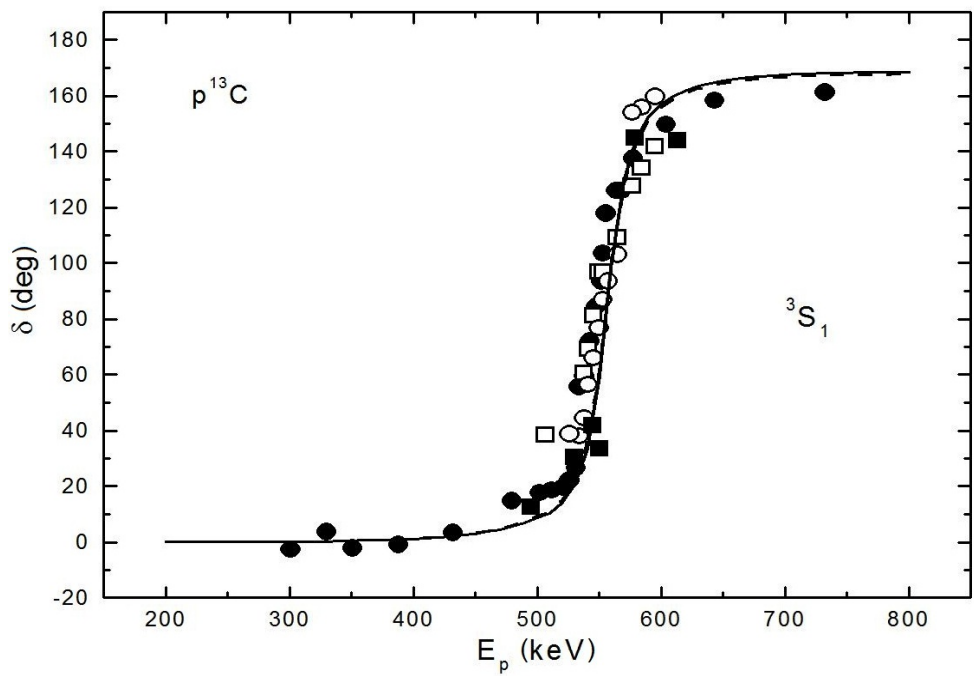

Fig. 8. The $^3S_1$ phase shift of the p$^{13}$C elastic scattering at astrophysical energies. Points: ●, ○, ■ and □ – our phase shift analysis based on the data from Ref. 85, 86. Lines: solid – the phase shift calculations with potential from Eq. (50) given in the text, dashed – calculations with potential from Eq. (51).

The potential with the FS of the $^3P_1$ bound state should represent correctly the binding energy of $^{14}$N with $J^PT = 1^+0$ in the p$^{13}$C channel[82] at -7.55063 MeV as well as describe reasonably the mean square charge radius of $^{14}$N, which has the experimental value 2.560(11) fm.[82] As a result, the following parameters were obtained:

$$V_{GS} = 1277.853205 \text{ MeV} \quad \text{and} \quad \alpha_{GS} = 1.5 \text{ fm}^{-2}. \tag{52}$$

The potential gives the binding energy equals -7.550630 MeV and the mean square charge radius $R_{ch} = 2.38$ fm. The values 0.8768(69) fm and 2.4628(39) fm were used as the radii of the proton and $^{13}$C, respectively.[13,82]

Another variant of the potential of the $^3P_1$ ground state of $^{14}$N defined as binary p$^{13}$C system

$$V_{GS} = 1679.445025 \text{ MeV} \quad \text{and} \quad \alpha_{GS} = 2.0 \text{ fm}^{-2} \tag{53}$$

leads to the binding energy -7.550630 MeV and also gives slightly understated mean square charge radius $R_{ch} = 2.36$ fm.

Note that such a high accuracy of the potential parameters is required for providing of the correct description of nuclear binding energy, up to $10^{-6}$ MeV.

### 7.2. *Radiative capture reaction*

In this work, we will continue the study of the astrophysical S-factors for reactions with light nuclei and will examine on the proton capture on $^{13}$C at astrophysical energies. This process is a part of the CNO thermonuclear cycle, which gives the essential contribution into the energy release of thermonuclear reactions,[1-3,5,89] leading to the burning of the Sun and stars of our Universe.[4] The existent experimental data, of the astrophysical S-factor of the radiative proton capture on $^{13}$C (see Ref. 10), show the presence of the narrow, with the width about 23(1) keV, resonance at the energy 0.551(1) MeV (l.s.)[82] that leads to the S-factor's rise by two-three orders. Such form of the S-factor can be obtained due to the E1 transition, with the spin change $\Delta T = 1$, from the $^3S_1$ resonance scattering state at 0.55 MeV and moments $J^P T = 1^- 1$ to the $^3P_1$ triplet bound state of p$^{13}$C clusters, with the potentials like in Eqs. (52) or (53). This state corresponds the ground state of $^{14}$N with quantum numbers $J^P T = 1^+ 0$ in the p$^{13}$C channel, since $^{13}$C has the moments $J^P T = 1/2^- 1/2$.[82]

The S-factor calculations of the radiative proton capture on $^{13}$C to the ground state of $^{14}$N with the above-cited potentials, for the $^3P_1$ state and the $^3S_1$ resonance scattering wave at the energies below 0.8 MeV, are given in Fig. 9 by the solid and dashed lines. The experimental results were taken from the work Adelberger *et al.*,[90] where, evidently, the most recent investigations of this reaction are reported. In the figure - the solid line is the result for combination of the potentials from Eqs. (50) and (52) and the dashed line for Eqs. (51) and (53). The calculation results of S-factor are very close, since the fact that the phase shifts of two $^3S_1$ scattering potentials, which are shown in Fig. 8, practically coincide. The calculated astrophysical S-factor for the first set of the potentials given in Eqs. (50) and (52) has practically constant value equals 5.0(1) keV b at the energy range 30÷50 keV. The value 4.8(1) keV b was obtained for the second set of the potentials from Eqs. (51) and (53) at the same energies. The fixed up error is obtained by the averaging of the S-factor value over the above noted energy range.

If Eq. (30) is used for parametrization of the S-factors, then the solid line in Fig. 9 can be described, at the energy range 30-200 keV with the average $\chi^2=0.11$, by the parameters: $S_0=4.1339$ keV b and $S_1=0.01844$ keV b keV$^{-1}$.

If the quadratic parametrization (Eq. (31)) is used, then the next parameters were obtained: $S_0$=4.9186 eV b, $S_1$=-3.3635·10$^{-6}$ eV b keV$^{-1}$ and $S_2$=8.5159·10$^{-5}$ eV b keV$^{-2}$, with the average $\chi^2$=0.00024 in the energy range 30-200 keV. Here, the 10% errors of the calculated $S$-factor values are used for the determination of $\chi^2$. If we will use the value 5.0 keV b for the $S$-factor at 10 keV, the next parameters will be obtained: $S_0$=4.9301 eV b, $S_1$=-3.3635·10$^{-6}$ eV b keV$^{-1}$ and $S_2$=8.4640·10$^{-5}$ eV b keV$^{-2}$ with the average $\chi^2$=0.0012 in the energy range 10-200 keV.

At the second method of parametrization, considered above, we obtain: $S_0$=5.0 keV b, $S_1$=-3.3635·10$^{-6}$ keV b keV$^{-1}$ and $S_2$=8.1472·10$^{-5}$ keV b keV$^{-2}$ with the average $\chi^2$=0.0078 in the energy range 10-200 keV.

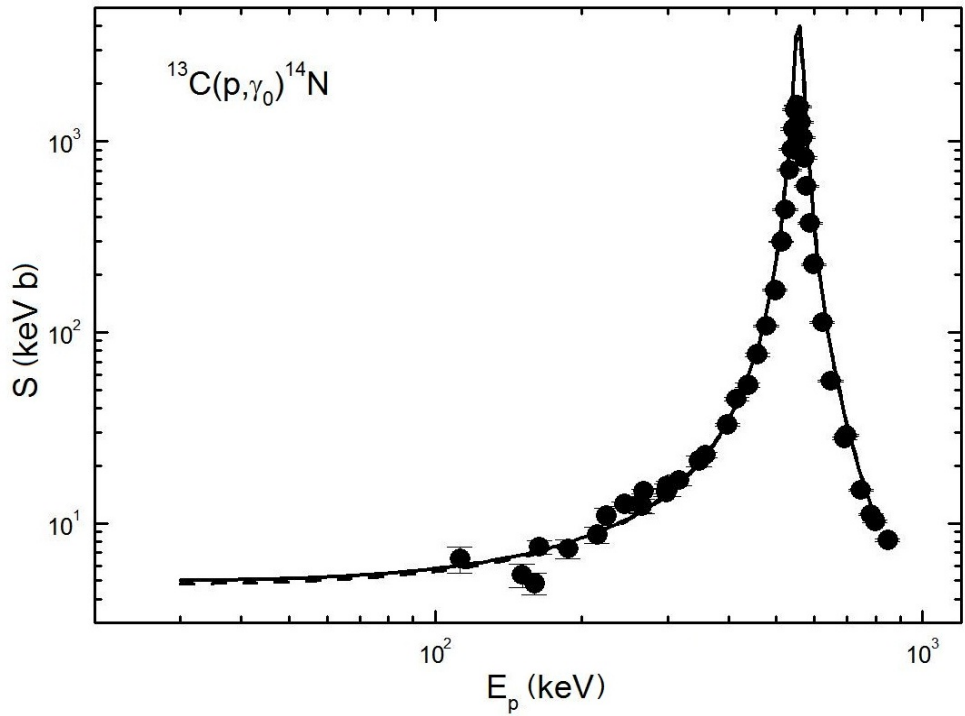

Fig. 9. The astrophysical $S$-factor of the radiative proton capture on $^{13}$C at low energies. Experimental points (●) are taken from King *et al.*[90] Lines: solid – our calculation with the first set of potentials from Eqs. (50) and (52), dashed – the calculation for the second set of potentials from Eqs. (51) and (53).

The known extrapolations of the measured $S$-factor to zero energy for the transitions to the ground state of $^{14}$N lead to the following values: 5.25 keV b;[90] it was obtained 5.16±72 keV b, on the basis of results from King *et al.*[90] in work Mukhamedzhanov *et al.*;[91] the value 5.36±71 keV b was found by Mukhamedzhanov *et al.*;[92] the value 5.5 keV b was suggested in measurements;[93] and the value 5.06 keV b, for the $S$-factor at zero energy, was given in the work Artemov *et al.*[94]

That is to say, the completely acceptable results in the description of the astrophysical $S$-factor of the radiative proton capture on $^{13}$C at the energy range from 30 to 800 keV are obtained for both considered variants of the p$^{13}$C intercluster potentials. The $S$-factor value is practically constant at energies 30÷50 keV, thereby determines its value for the extrapolation to zero energy. The obtained results confirm[88] the fact that to parametrize reliable intercluster scattering potentials and obtain, on their basis, the characteristics of nuclei and nuclear processes is possible only with the enough accurate determination of the experimental data of the elastic cluster scattering phase shifts. The parameters for the BS potentials are fixed by the nuclear characteristics quite definitely. Unfortunately, at present time, the scattering phase shifts are determined with largish errors, for the majority of lightest nuclei; sometimes reach to 20-30%.[65] In this connection, in terms of future usage of the PCM, the problem of accuracy increasing for the experimental data of the elastic scattering of light atomic nuclei at low and astrophysical energies, which are needed for the construction of the intercluster potentials, is quite urgent until the present time.

## 8. Conclusion

The description of the *S*-factor's behavior in all of the considered systems at low energies may be viewed as a certain testimony in favor of the potential approach in the cluster model. The inter-cluster interactions with FS, the structure of which is determined by the orbital state classification according to the Young schemes, are constructed on the basis of the phase shifts of the elastic scattering of clusters, and each partial wave is described, for example, by its potential of Gaussian form with certain parameters.

Separation of the general interaction into the partial waves allows specifying its structure, while the classification of the orbital states according to the Young schemes allows identifying the presence and the number of forbidden states. It gives the possibility to find the number of nodes of the cluster relative motion WF of the GS of nuclei and leads to a definite depth of interaction, which helps to avoid the discrete ambiguity of the potential depth as it is the case in the optical model.

The form of each partial phase shift of scattering can be correctly described at the certain width of such a potential only and it saves us from the continuous ambiguity, which also takes place in the conventional optical model. Consequently, all parameters of such a potential are fixed quite uniquely, and the "pure" (according to Young schemes) interaction component allows describing the basic characteristics of the bound state of the lightest clusters generally correctly, which is realized in the light atomic nuclei with a high probability. The requirement of a correct description of the BS characteristics, in partial waves where they exist, is an additional criterion for the determination of the intercluster potential parameters.

The formalism developed for obtaining the intercluster potentials was applied here for description of nuclear photocapture reactions in considered systems. The operator of electromagnetic transition for radiative capture processes, as opposed to the one for nuclear reactions due to strong interaction, is well known. Moreover, the interaction in the final state is absent in photocapture reactions and the interaction in the initial state is taken into account rather correctly based on the developed potential approach.

The calculations of the *S*-factor of the radiative proton capture on $^2$H at energies down to 10 keV in the framework of this potential cluster model were carried out at the times when experimental data only for the energies above 150÷200 keV were well-known, and their results are in a good agreement with the experimental data in the range from 50 keV to 150÷200 keV that became available much later. That is to say, the prediction of *S*-factor's behavior in $^2$H(p,γ)$^3$He at this energy range was made.

However, as it has been shown above, the reliable results for the intercluster potentials and, consequently, for the characteristics of nuclear processes can be obtained only if the phase shifts of elastic scattering are accurately determined in the experiment. Unfortunately, at present time for the majority of the lightest nuclear systems the elastic scattering phase shifts are found with significant errors reaching sometimes 20-30%. This makes the construction of the exact potentials of intercluster interaction very difficult and, finally, leads to significant ambiguities in the final results obtained in the potential cluster model.

In this connection, it is very urgent to raise the accuracy of experimental measurements of elastic scattering of light atomic nuclei at astrophysical energies and to perform a more accurate phase shift analysis. The increase in the accuracy and the use of this semi-microscopic and semi-phenomenological model will allow making more definite conclusions regarding the mechanisms and conditions of thermonuclear reactions as well as understanding their nature in general.

**Acknowledgments**

Finally, the authors would like to express their deepest gratitude to Prof. A. Mukhamedzhanov (Cyclotron Institute, Texas A&M University), Prof. Yu. N. Uzikov (JIRN, Dubna, Russia), Prof. B. S. Ishkhanov and Prof. L. D. Blokhintsev (Moscow State University, Moscow, Russia) for the very important discussions of some parts of the work.

This work was supported in part by the V. G. Fessenkov Astrophysical Institute NCSRT NSA RK with the help of Grant Program of Fundamental Research of the Ministry of Education and Science of the Republic of Kazakhstan.